\documentclass{aa}

\usepackage{txfonts}
\usepackage[dvips]{graphicx}
\usepackage{amsfonts}
\usepackage{amssymb}
\usepackage{natbib}
\usepackage{morefloats}
\usepackage{color,ulem}
\usepackage{lscape}
\usepackage{longtable}
\usepackage{rotating}
\bibpunct{(}{)}{;}{a}{}{,}

\newcommand{\GD}{$\gamma$~Dor}
\newcommand{\DSct}{$\delta$~Sct}
\newcommand{\cd}{d$^{-1}$}
\newcommand{\mhz}{$\mu$Hz}
\newcommand{\te}{$T_{\rm eff}$}
\newcommand{\logg}{$\log{g}$}
\newcommand{\vsini}{$v\sin{i}$}
\newcommand{\kms}{km\,s$^{-1}$}

\begin{document}
\title{Detection of a large sample of $\gamma$~Dor stars from $Kepler$ space photometry and high-resolution ground-based spectroscopy.\thanks{Based on data gathered with NASA Discovery mission $Kepler$ and spectra obtained with the HERMES spectrograph, installed at the Mercator
    Telescope, operated on the island of La Palma by the Flemish Community, at
    the Spanish Observatorio del Roque de los Muchachos of the Instituto de
    Astrof\'{\i}sica de Canarias and supported by the Fund for Scientific
    Research of Flanders (FWO), Belgium, the Research Council of KU Leuven,
    Belgium, the Fonds National de la Recherche Scientifique (F.R.S.--FNRS),
    Belgium, the Royal Observatory of Belgium, the Observatoire de Gen\`eve,
    Switzerland and the Th\"uringer Landessternwarte Tautenburg, Germany.}$^,$\thanks{Tables A.1 and B.1 are only available in electronic form at the CDS via anonymous ftp to cdsarc.u-strasbg.fr (130.79.128.5) or via http://cdsweb.u-strasbg.fr/cgi-bin/qcat?J/A+A/}}
\author{A. Tkachenko\inst{1,}\thanks{Postdoctoral Fellow of the Fund for Scientific Research (FWO), Flanders, Belgium} \and C. Aerts\inst{1,2} \and A. Yakushechkin\inst{3} \and J. Debosscher\inst{1} \and P. Degroote\inst{1}\thanks{Postdoctoral Fellow of the Fund for Scientific Research (FWO), Flanders, Belgium} \and S. Bloemen\inst{1} \and P.~I.\,P\'{a}pics\inst{1} \and B.~L.\ de Vries\inst{1}\thanks{Aspirant Fellow of the Fund for Scientific Research (FWO), Flanders, Belgium} \and R.\ Lombaert\inst{1} \and M.\ Hrudkova\inst{4} \and Y.\ Fr\'{e}mat\inst{5} \and G.\ Raskin\inst{1} \and H.\ Van Winckel\inst{1}}
\institute{Instituut voor Sterrenkunde, KU Leuven, Celestijnenlaan
200D, B-3001 Leuven, Belgium\\ \email{andrew@ster.kuleuven.be} \and
Department of Astrophysics, IMAPP, University of Nijmegen, PO Box
9010, 6500 GL Nijmegen, The Netherlands \and Tavrian National
University, Department of Astronomy, Simferopol, Ukraine \and Isaac
Newton Group of Telescopes, Apartado de Correos 321, E-387 00 Santa
Cruz de la Palma, Canary Islands, Spain \and Royal Observatory of
Belgium, 3 Avenue Circulaire 1180 Brussel, Belgium}
\titlerunning{Selection of a sample of 69 $\gamma$~Dor stars for asteroseismology}
\date{Received date; accepted date}
\abstract{The launches of the MOST, CoRoT, and $Kepler$ missions
opened up a new era in asteroseismology, the study of stellar
interiors via interpretation of pulsation patterns observed at the
surfaces of large groups of stars. These space-missions deliver a
huge amount of high-quality photometric data suitable to study
numerous pulsating stars.}{Our ultimate goal is a detection and
analysis of an extended sample of \GD-type pulsating stars with the
aim to search for observational evidence of non-uniform period
spacings and rotational splittings of gravity modes in main-sequence
stars typically twice as massive as the Sun. This kind of diagnostic
can be used to deduce the internal rotation law and to estimate the
amount of rotational mixing in the near core regions.}{We applied an
automated supervised photometric classification method to select a
sample of 69 Gamma Doradus (\GD) candidate stars. We used an
advanced method to extract the $Kepler$ light curves from the pixel
data information using custom masks. For 36 of the stars, we
obtained high-resolution spectroscopy with the HERMES spectrograph
installed at the Mercator telescope. The spectroscopic data are
analysed to determine the fundamental parameters like \te, \logg,
\vsini, and [M/H].}{We find that all stars for which spectroscopic
estimates of \te\ and \logg\ are available fall into the region of
the HR diagram where the \GD\ and \DSct\ instability strips overlap.
The stars cluster in a 700~K window in effective temperature, \logg\
measurements suggest luminosity class IV-V, i.e. sub-giant or
main-sequence stars. From the $Kepler$ photometry, we identify 45
\GD-type pulsators, 14 \GD/\DSct\ hybrids, and 10 stars which are
classified as ``possibly \GD/\DSct\ hybrid pulsators''. We find a
clear correlation between the spectroscopically derived \vsini\ and
the frequencies of independent pulsation modes.}{We have shown that
our photometric classification based on the light curve morphology
and colour information is very robust. The results of spectroscopic
classification are in perfect agreement with the photometric
classification. We show that the detected correlation between
\vsini\ and frequencies has nothing to do with rotational modulation
of the stars but is related to their stellar pulsations. Our sample
and frequency determinations offer a good starting point for seismic
modelling of slow to moderately rotating \GD\ stars.}
 \maketitle

\section{Introduction}

Asteroseismology is a powerful tool for diagnostics of deep stellar
interiors not accessible through observations otherwise. Detection
of as many oscillation modes as possible that propagate from the
inner layers of a star all the way up to the surface is a clue
towards better understanding of stellar structure at different
evolutionary stages via asteroseismic methods. The recent launches
of the MOST \citep[Microvariability and Oscillations of
STars,][]{Walker2003}, CoRoT \citep[Convection Rotation and
Planetary Transits,][]{Auvergne2009} and $Kepler$
\citep{Gilliland2010a} space missions delivering high-quality
photometric data at a micro-magnitude precision led to a discovery
of numerous oscillating stars with rich pulsation spectra and opened
up a new era in asteroseismology.

Besides acoustic waves propagating throughout the stars and having
the pressure force as their dominant restoring force (p-modes), some
stars pulsate also in the so-called g-modes for which gravity (or
buoyancy) is the dominant restoring force. Given that gravity modes
have high amplitudes deep inside stars, they allow to study
properties of the stellar core and nearby regions far better than
acoustic modes.

\begin{table} \tabcolsep 3.0mm\caption{\small Influence of the pixel
mask choice on the frequency analysis for KIC\,08364249. The
uncertainty in frequency is given by the Rayleigh limit (1/$T$)
which amounts 0.0011~\cd\ (0.0127~\mhz) in this case. The barred
values indicate frequencies which are not present when using our
custom mask to produce the light curve but are detected when using
the standard mask data. The frequencies are sorted according to the
amplitude.}\label{Table: Mask comparison}
\begin{tabular}{llllll}
\hline\hline
F$_{\rm i}$\rule{0pt}{9pt} & \multicolumn{2}{c}{Frequency} & F$_{\rm i}$ & \multicolumn{2}{c}{Frequency}\\
& \multicolumn{1}{c}{\cd} & \multicolumn{1}{c}{\mhz} & & \multicolumn{1}{c}{\cd} & \multicolumn{1}{c}{\mhz}\\
\hline
F$_{ 1}$\rule{0pt}{9pt}& 1.8693 &   21.6282 & F$_{ 33}$& 3.8555 &   44.6083 \\
F$_{ 2}$\rule{0pt}{9pt}& 2.0095 &   23.2498 & F$_{ 34}$& \sout{0.0318} &    \sout{0.3684} \\
F$_{ 3}$\rule{0pt}{9pt}& 1.9862 &   22.9802 & F$_{ 35}$& \sout{0.0164} &    \sout{0.1895} \\
F$_{ 4}$\rule{0pt}{9pt}& 1.8893 &   21.8597 & F$_{ 36}$& 3.9690 &   45.9208 \\
F$_{  5}$\rule{0pt}{9pt}& 1.8633 &   21.5585 & F$_{ 37}$& 1.7565 &   20.3222 \\
F$_{  6}$\rule{0pt}{9pt}& 1.9194 &   22.2069 & F$_{ 38}$& \sout{0.0370} &    \sout{0.4276} \\
F$_{  7}$\rule{0pt}{9pt}& \sout{0.0202} &    \sout{0.2342} & F$_{ 39}$& 3.8788 &   44.8779 \\
F$_{  8}$\rule{0pt}{9pt}& 1.7523 &   20.2736 & F$_{ 40}$& 1.8684 &   21.6177 \\
F$_{  9}$\rule{0pt}{9pt}& 1.9040 &   22.0293 & F$_{ 41}$& \sout{0.0626} &    \sout{0.7248} \\
F$_{ 10}$\rule{0pt}{9pt}& \sout{0.0123} &    \sout{0.1422} & F$_{ 42}$& 1.7759 &   20.5471 \\
F$_{ 11}$\rule{0pt}{9pt}& 1.9276 &   22.3029 & F$_{ 43}$& 3.7388 &   43.2576 \\
F$_{ 12}$\rule{0pt}{9pt}& 0.0224 &    0.2592 & F$_{ 44}$& 1.9380 &   22.4226 \\
F$_{ 13}$\rule{0pt}{9pt}& \sout{0.0182} &    \sout{0.2106} & F$_{ 45}$& \sout{0.0863} &    \sout{0.9984} \\
F$_{ 14}$\rule{0pt}{9pt}& 1.7293 &   20.0079 & F$_{ 46}$& \sout{0.0079} &    \sout{0.0909} \\
F$_{ 15}$\rule{0pt}{9pt}& \sout{0.0144} &    \sout{0.1672} & F$_{ 47}$& 3.9957 &   46.2299 \\
F$_{ 16}$\rule{0pt}{9pt}& 1.8701 &   21.6374 & F$_{ 48}$& 1.9855 &   22.9723 \\
F$_{ 17}$\rule{0pt}{9pt}& 3.7617 &   43.5233 & F$_{ 49}$& \sout{0.0283} &    \sout{0.3276} \\
F$_{ 18}$\rule{0pt}{9pt}& 1.8463 &   21.3612 & F$_{ 50}$& 4.0189 &   46.4982 \\
F$_{ 19}$\rule{0pt}{9pt}& 1.8314 &   21.1890 & F$_{ 51}$& 4.0152 &   46.4561 \\
F$_{ 20}$\rule{0pt}{9pt}& \sout{0.0102} &    \sout{0.1185} & F$_{ 52}$& 0.1203 &    1.3916 \\
F$_{ 21}$\rule{0pt}{9pt}& 0.1402 &    1.6217 & F$_{ 53}$& 1.9365 &   22.4055 \\
F$_{ 22}$\rule{0pt}{9pt}& 2.0642 &   23.8824 & F$_{ 54}$& 2.0958 &   24.2480 \\
F$_{ 23}$\rule{0pt}{9pt}& 3.9474 &   45.6710 & F$_{ 55}$& 1.8825 &   21.7808 \\
F$_{ 24}$\rule{0pt}{9pt}& \sout{0.0260} &    \sout{0.3013} & F$_{ 56}$& 1.9114 &   22.1148 \\
F$_{ 25}$\rule{0pt}{9pt}& \sout{0.0297} &    \sout{0.3434} & F$_{ 57}$& 1.9454 &   22.5081 \\
F$_{ 26}$\rule{0pt}{9pt}& 1.8577 &   21.4941 & F$_{ 58}$& 3.8523 &   44.5715 \\
F$_{ 27}$\rule{0pt}{9pt}& 2.0796 &   24.0613 & F$_{ 59}$& 1.8885 &   21.8505 \\
F$_{ 28}$\rule{0pt}{9pt}& 1.8926 &   21.8978 & F$_{ 60}$& 1.7235 &   19.9408 \\
F$_{ 29}$\rule{0pt}{9pt}& 0.1168 &    1.3508 & F$_{ 61}$& 1.7716 &   20.4972 \\
F$_{ 30}$\rule{0pt}{9pt}& 1.8754 &   21.6979 & F$_{ 62}$& 3.9916 &   46.1826 \\
F$_{ 31}$\rule{0pt}{9pt}& \sout{0.0648} &    \sout{0.7498} & F$_{ 63}$& 1.9871 &   22.9907 \\
F$_{ 32}$\rule{0pt}{9pt}& 2.0495 &   23.7128 & F$_{ 64}$& 2.2176 &   25.6579 \\
\hline
\end{tabular}
\end{table}

In this paper, we focus on the class of main-sequence, non-radially
pulsating stars named after the prototype star Gamma Doradus (\GD,
hereafter), whose multiperiodic variable nature was first reported
by \citet{Cousins1992}. \GD\ stars are assumed to pulsate in
high-order, low-degree gravity modes driven by the flux blocking
mechanism near the base of their convective zones
\citep{Guzik2000,Dupret2005}. These stars have masses ranging from
1.5 to 1.8~M$_{\odot}$ \citep{Aerts2010} and A7--F5 spectral types
\citep{Kaye1999}. They are usually multiperiodic pulsators
exhibiting both low-amplitude photometric and spectroscopic
variability with periods between 0.5 and 3 days \citep{Kaye1999}.
\GD\ stars are grouped in a region close to the red edge of the
classical instability strip in the Hertzsprung–-Russell (HR)
diagram. The theoretical \GD\ instability strip overlaps with the
region where the Delta Scuti (\DSct, hereafter) stars pulsating in
low-order p-modes are located. Pulsators in the overlapping region
are expected to show both high-order g-modes probing the core and
low-order p- and g-modes probing the outer layers. Such {\it hybrid
pulsators} are among the most interesting targets for asteroseismic
diagnostics as the co-existence of the two types of oscillation
modes yields potential of constraining the whole interior of the
star, from the core to the outer atmosphere.

The first-order asymptotic approximation for non-radial pulsations
developed by \citet{Tassoul1980} shows that, in the case of a model
representative of a typical \GD\ pulsator, the periods of
high-order, low-degree g-modes are equally spaced. However, the
steep chemical composition gradient left by the shrinking core in
stars with masses above $\sim$1.5~M$_{\odot}$ causes sharp features
in the Brunt-V\"{a}is\"{a}l\"{a} frequency to occur, which in turn
determines how the gravity modes are trapped in the stellar
interiors. The trapping also leads to the deviation from uniform
period spacing of the g-modes of consecutive radial order $n$
\citep{Miglio2008}. Moreover, the rotational splitting of the
g-modes provides a unique opportunity to deduce the internal
rotation law and to estimate the amount of rotational mixing that
may be partly responsible for the chemically inhomogeneous layer
exterior to the core. This type of diagnostic is not available for
stars pulsating purely in p-modes. Consequently, a non-uniform
period spacing together with the rotational splitting of gravity
modes represent a powerful tool for the diagnostics of the
properties of the inner core and surrounding layers.

This idea was worked out in full detail for white dwarfs by
\citet{Brassard1992} and applied to such type of objects by, e.g.,
\citet{Winget1991} and \citet{Costa2008}. \citet{Miglio2008}
developed similar methodology for gravity modes of main-sequence
pulsators. Recently, \citet{Bouabid2013} studied the effects of the
Coriolis force and of diffusive mixing on high-order g-modes in \GD\
stars, adopting the traditional approximation. The authors conclude
that rotation has no significant influence on instability strips but
does allow to fill the gap between \DSct-type p- and \GD-type
g-modes in the frequency spectrum of a pulsating star by shifting
the g-mode frequencies to higher values. \citet{Bouabid2013} also
conclude that the deviations from a uniform period spacing is no
longer periodically oscillating around a constant value, but as far
as we are aware this theoretical description was not yet applied to
data.

A first concrete application of detected period spacings to
main-sequence stars was done by \citet{Degroote2010}, based on CoRoT
data of the Slowly Pulsating B (hereafter called SPB) star
HD\,50230. Among some hundreds statistically significant
frequencies, the authors identify eight high-amplitude peaks in the
g-mode regime showing a clear periodic deviation from the mean
period spacing of 9418~s. They link this deviation with the chemical
gradient left by the shrinking core and constrain the location of
the transition zone to be at about 10 per cent of the radius.
Additionally, the authors suggest an extra-mixing around the
convective core to occur. Another indication of a non-uniform period
spacing was reported by \citet{Papics2012b} for the SPB/$\beta$\,Cep
hybrid pulsator HD\,43317. The authors detect two series of ten and
seven components in the g-mode regime showing mean period spacings
of 6339 and 6380~s, respectively. Given that the two values of the
period spacing are almost the same, the authors conclude that both
series belong to the same value of degree $l$ but have different
azimuthal numbers $m$. However, they do not present any physical
interpretation of the observed period spacings, claiming that the
detected modes are in the gravito-inertial regime, and thus full
computations with respect to the rotation of the star, rather than
approximations based on a perturbative approach are required.

The class of \GD-type pulsators should contain both stars with a
shrinking and expanding core during their main-sequence evolution,
and seismic probing of the core holds the potential to discriminate
between those two scenarios. Based on the analysis of almost 11\,500
light curves gathered with the CoRoT space mission,
\citet{Hereter2010} confidently detect 34 \GD\ stars and 25
\GD/\DSct\ hybrid pulsators with about 50 stars still being
candidates of either of the class. For five out of 34 \GD\ stars,
the authors detect regular patterns of period spacings of high-order
gravity modes ranging from 650 to 2\,400~s. \citet{Hareter2012}
extends the sample up to 95 and 54 \GD\ and hybrid pulsators,
respectively, and attempts to interpret the period spacings from the
previous study. The author finds that the observed spacings are too
short to be due to high-order $l=1$ modes and concludes that these
spacings are either coincidental due to the rather poor Rayleigh
limit of CoRoT data or need some other explanation. Recently,
\citet{Maceroni2013} studied an eclipsing binary, CoRoT 102918586,
with a \GD\ pulsating component based on space photometry and
ground-based spectroscopy. The authors detect a nearly equidistant
frequency spacing of about 0.05~\cd\ (0.58~\mhz) which was found to
be too small to be due to the stellar rotation. The corresponding
mean period spacing of 3110~s was found to be consistent with the
theoretical spacings of $l=1$ high-order g-modes based on the models
computed without convective overshooting. Neither of the two studies
present a detailed modelling of the deviation from the constant
period spacings, however, and both rely on comparison with
theoretically predicted mean values of the period spacings for an
F-type star. This encourages us to start this work, with the aim of
studying in more detail the frequency content of \GD\ stars observed
in the $Kepler$ data, which have a much more suitable Rayleigh limit
to unravel the different structures observed in their amplitude
spectra.

Our ultimate goal is to deduce with high precision the extent of the
stellar core and mixing processes near the core for \GD-like stars
burning hydrogen in their cores. This requires the selection of a
large enough sample of slowly to moderately rotating \GD\ stars and
this paper is devoted to this purpose. Both photometric and
spectroscopic data have been acquired and are described together
with the data reduction process in Sect.~\ref{Section:
Observations}. The sample of stars is introduced in
Sect.~\ref{Section: The target_sample}. Photometric characterization
of the sample is given in Sect.~\ref{Section: Frequency analysis},
the results of the spectroscopic analysis of all stars for which
spectra are available are outlined in Sect.~\ref{Section:
Spectroscopic analysis}. The conclusions and an outlook for future
research are given in Sect.~\ref{Section: Conclusions}.

\begin{table}
\caption{Journal of spectroscopic observations. N gives the number
of obtained spectra, $V$ -- visual magnitude. All spectra have been
taken in 2011. The stars are sorted according to the KIC
number.}\label{Table: Observations}
\begin{tabular}{lllll}
\hline\hline
\multicolumn{1}{c}{KIC\rule{0pt}{9pt}} & Designation & \multicolumn{1}{c}{$V$} & N & Observed\\
\hline
02710594\rule{0pt}{11pt} & TYC\,3134--1158--1 & 11.8 & 4 & May -- August\vspace{0.3mm} \\
03448365 & BD+38\,3623 & \,\,\,9.9 & 2 & May -- June\vspace{0.3mm} \\
04547348 & TYC\,3124--1108--1 & 11.4 & 4 & May -- August\vspace{0.3mm} \\
04749989 & TYC\,3139--151--1 & \,\,\,9.6 & 2 & May -- June\vspace{0.3mm} \\
04757184 & TYC\,3139--499--1 & 11.7 & 4 & May -- August\vspace{0.3mm} \\
05114382 & TYC\,3140--1590--1 & 11.6 & 4 & July -- August\vspace{0.3mm} \\
05522154 & TYC\,3125--2566--1 & 10.4 & 3 & May -- June\vspace{0.3mm} \\
05708550 & TYC\,3139--428--1 & 11.9 & 4 & July -- August\vspace{0.3mm} \\
06185513 & TYC\,3127--2073--1 & 11.9 & 4 & May\vspace{0.3mm} \\
06425437 & TYC\,3128--1004--1 & 11.5 & 3 & May\vspace{0.3mm} \\
06468146 & HD\,226446 & 10.0 & 2 & July\vspace{0.3mm} \\
06678174 & TYC\,3129--3189--1 & 11.7 & 4 & May -- August\vspace{0.3mm} \\
06935014 & TYC\,3128--1500--1 & 10.9 & 5 & May -- June\vspace{0.3mm} \\
07023122 & BD+42\,3281 & 10.8 & 4 & May -- July\vspace{0.3mm} \\
07365537 & BD+42\,3365 & \,\,\,9.2 & 3 & May\vspace{0.3mm} \\
07380501 & TYC\,3144--1294--1 & 12.5 & 4 & July -- August\vspace{0.3mm} \\
07867348 & TYC\,3130--1278--1 & 11.0 & 4 & May\vspace{0.3mm} \\
08364249 & ----- & 11.9$^*$ & 4 & July -- August\vspace{0.3mm} \\
08375138 & BD+44\,3210 & 11.0 & 4 & May -- August\vspace{0.3mm} \\
08378079 & TYC\,3148--1427--1 & 12.5 & 3 & August\vspace{0.3mm} \\
08611423 & TYC\,3132--580--1 & 11.6 & 4 & May -- June\vspace{0.3mm} \\
08645874 & HD\,188565 & \,\,\,9.9 & 2 & July\vspace{0.3mm} \\
09210943 & TYC\,3542--291--1 & 11.8 & 3 & May -- August\vspace{0.3mm} \\
09751996 & TYC\,3540--1251--1 & 11.0 & 4 & May\vspace{0.3mm} \\
10080943 & TYC\,3560--2433--1 & 11.7 & 4 & August\vspace{0.3mm} \\
10224094 & TYC\,3561--399--1 & 12.5 & 4 & July -- August\vspace{0.3mm} \\
11099031 & TYC\,3562--121--1 & 10.0 & 4 & July\vspace{0.3mm} \\
11294808 & TYC\,3551--405--1 & 11.7 & 4 & May -- August\vspace{0.3mm} \\
11721304 & TYC\,3565--1474--1 & 11.7 & 4 & July -- August\vspace{0.3mm} \\
11826272 & BD+49\,3115 & 10.2 & 3 & May -- August\vspace{0.3mm} \\
11907454 & TYC\,3550--369--1 & 11.7 & 4 & May -- August\vspace{0.3mm} \\
11917550 & TYC\,3564--346--1 & 11.1 & 4 & May -- August\vspace{0.3mm} \\
11920505 & TYC\,3564--2927--1 & \,\,\,9.9 & 2 & May\vspace{0.3mm} \\
12066947 & TYC\,3564--16--1 & 10.2 & 3 & May -- July\vspace{0.3mm} \\
12458189 & TYC\,3555--240--1 & 11.6 & 4 & May -- August\vspace{0.3mm} \\
12643786 & TYC\,3554--1916--1 & 11.6 & 4 & May -- August\vspace{0.3mm} \\
\hline \multicolumn{5}{l}{$^*$$Kepler$ magnitude\rule{0pt}{9pt}}
\end{tabular}
\end{table}

\section{Observations and data reduction}\label{Section:
Observations}

We base our analysis on high-quality photometric and spectroscopic
data. The light curves have been gathered by the $Kepler$ space
telescope and are of micro-magnitude precision. The data are
acquired in two different modes, the so-called $long$ (LC) and
$short$ (SC) cadences. In both cases exposure time is 6.54~s of
which 0.52~s is spent for the readout. The LC integrate over 270
single exposures giving a time-resolution of 29.42~min whereas one
data point in the SC mode contains 9 exposures corresponding to a
58.85-s time resolution. The data are released in quarters ($\sim$ 3
months of nearly continuous observations), the periods between
spacecraft rolls needed to keep its solar panels pointing towards
the Sun. More information on the characteristics of the SC and LC
data can be found in \citet{Gilliland2010b} and \citet{Jenkins2010},
respectively.

\begin{figure*}
\centering
\includegraphics[scale=1.0]{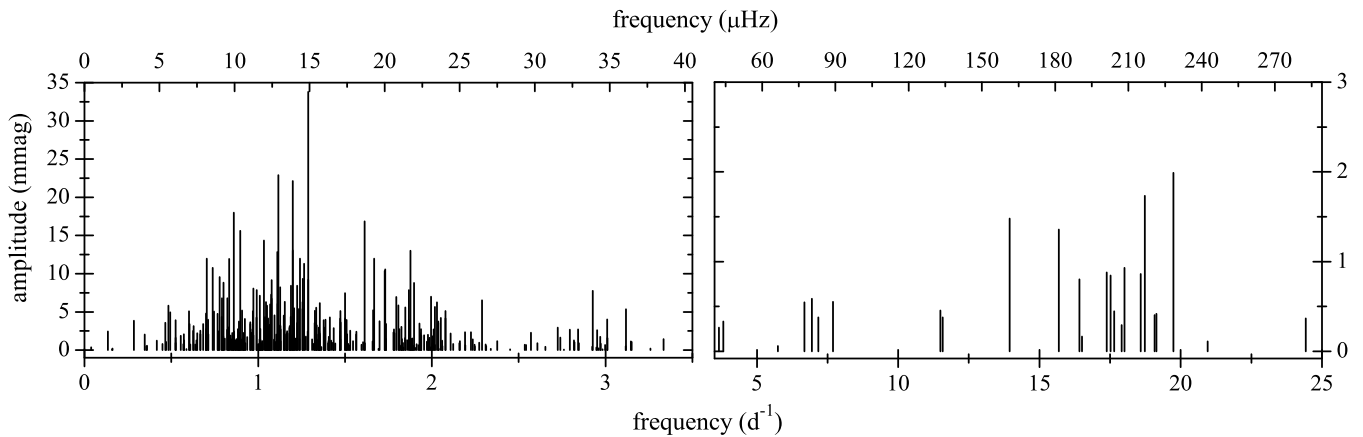}
\caption{\small The target sample representation. The plot
illustrates all frequencies we consider as independent ones for all
69 stars. Note the different Y-scales for the two panels.}
\label{Figure: Sample representation}
\end{figure*}

\begin{figure*}[t]
\centering
\includegraphics[scale=0.77]{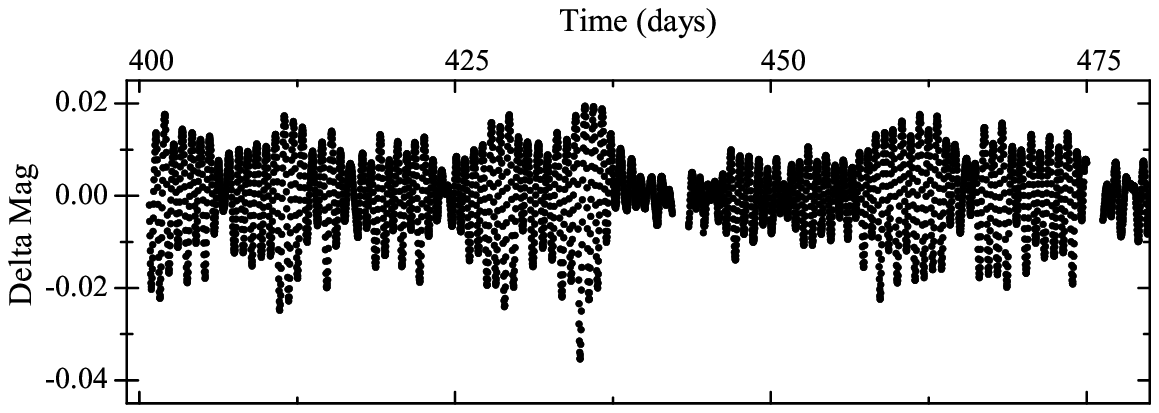}
\includegraphics[scale=0.77]{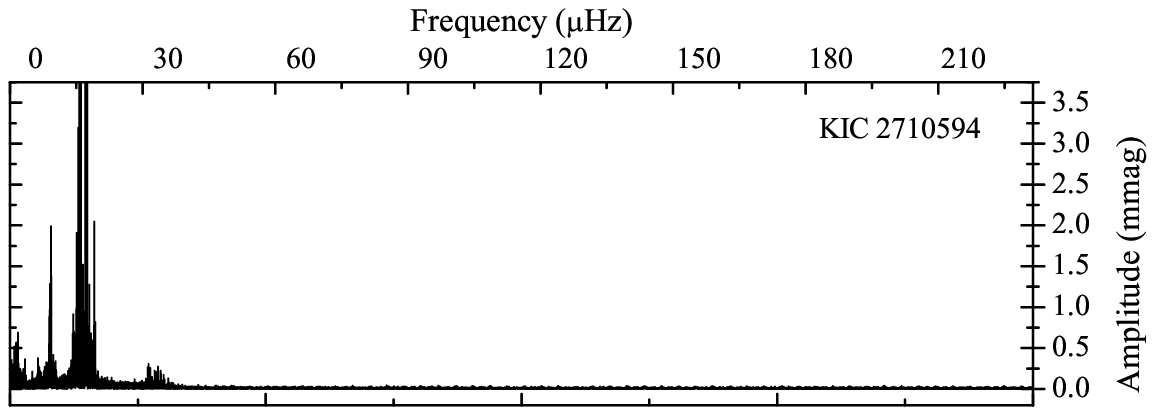}
\includegraphics[scale=0.77]{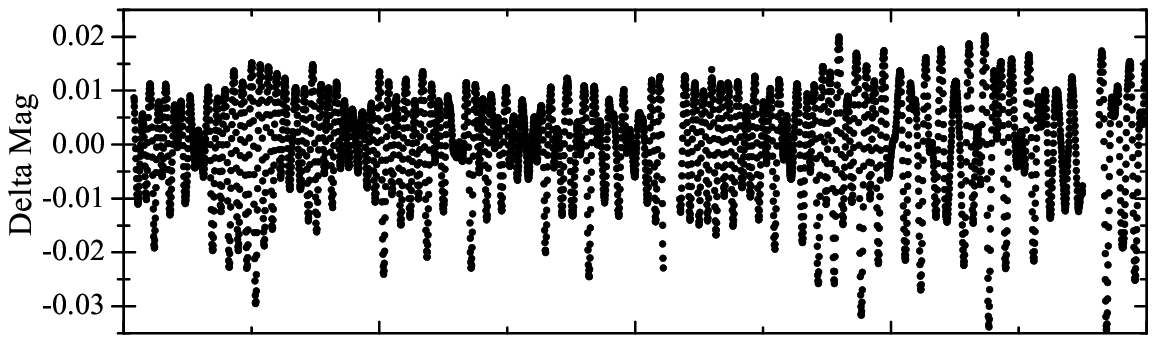}
\includegraphics[scale=0.77]{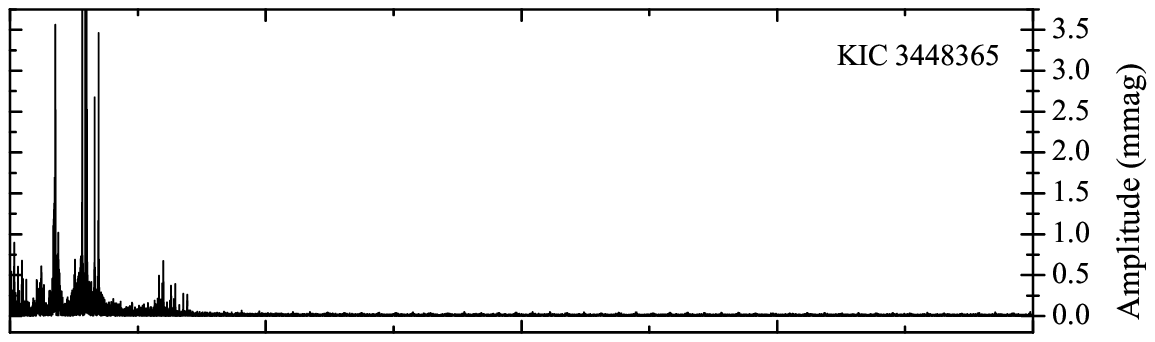}
\includegraphics[scale=0.77]{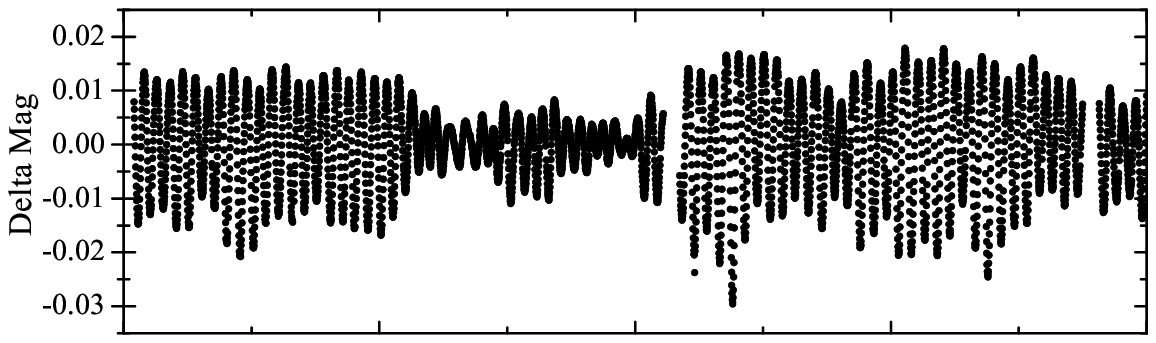}
\includegraphics[scale=0.77]{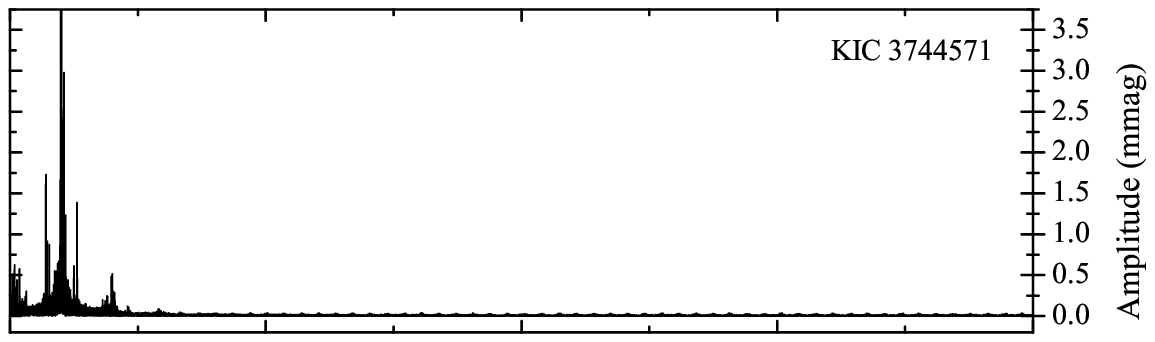}
\includegraphics[scale=0.77]{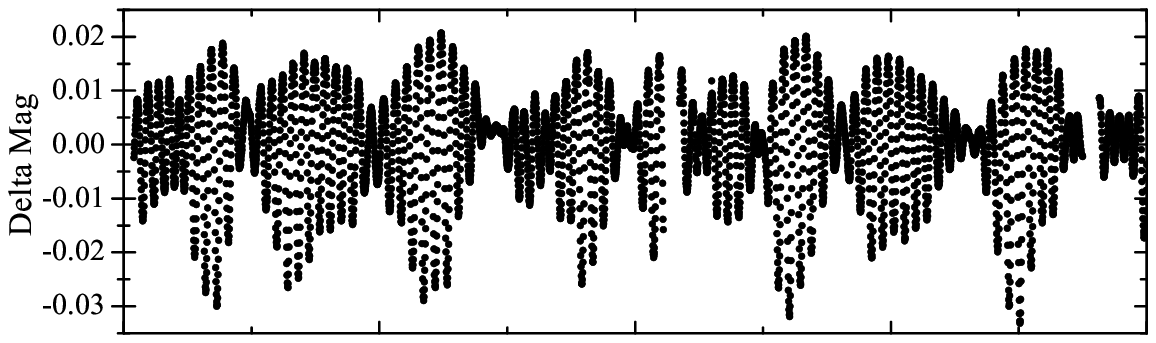}
\includegraphics[scale=0.77]{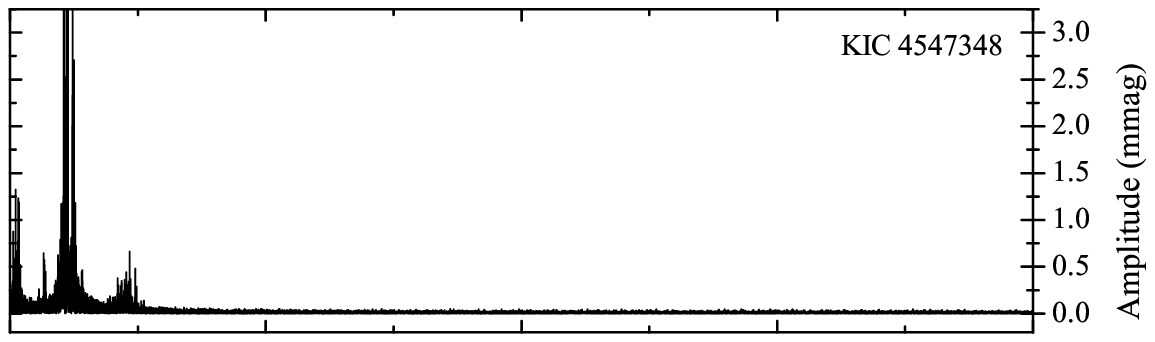}
\includegraphics[scale=0.77]{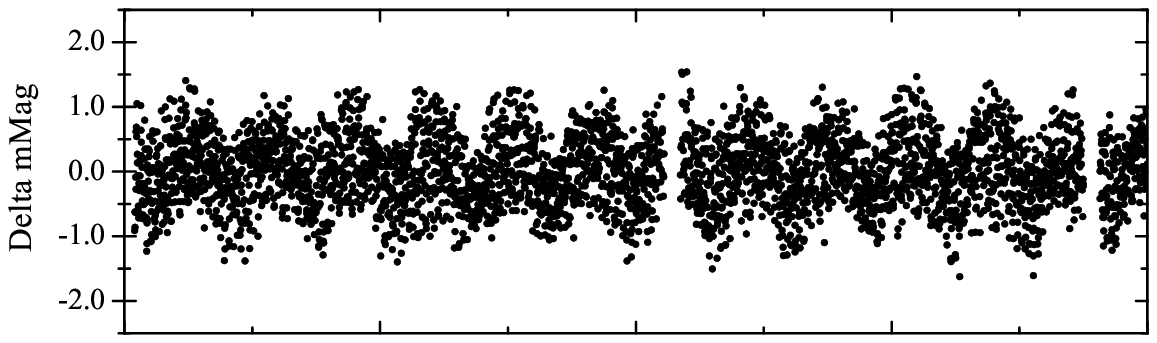}
\includegraphics[scale=0.77]{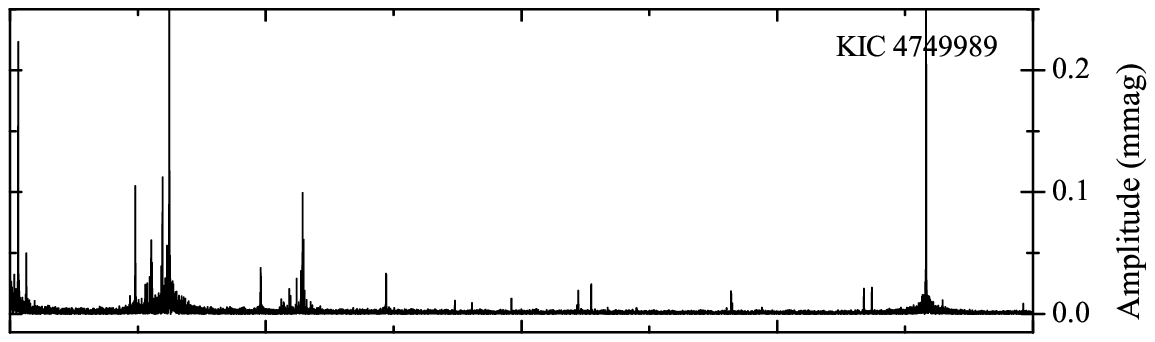}
\includegraphics[scale=0.77]{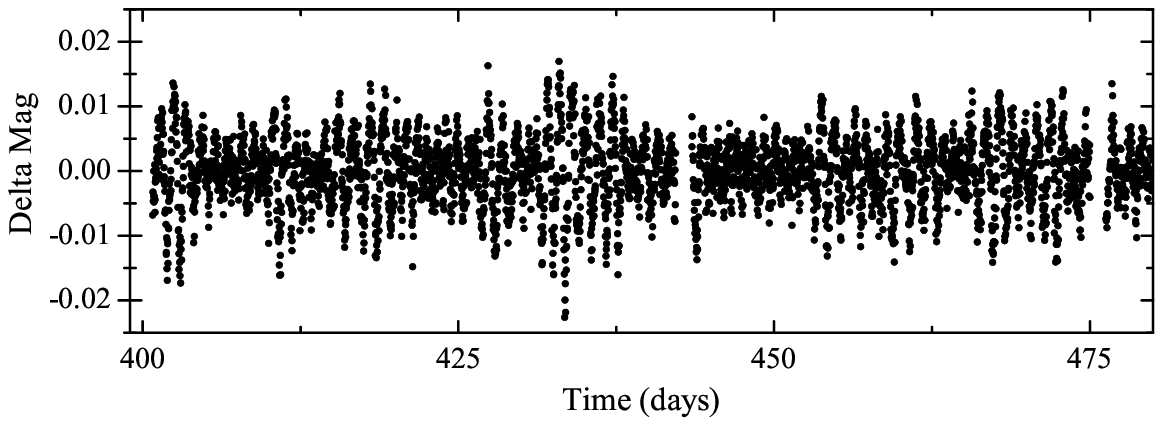}
\includegraphics[scale=0.77]{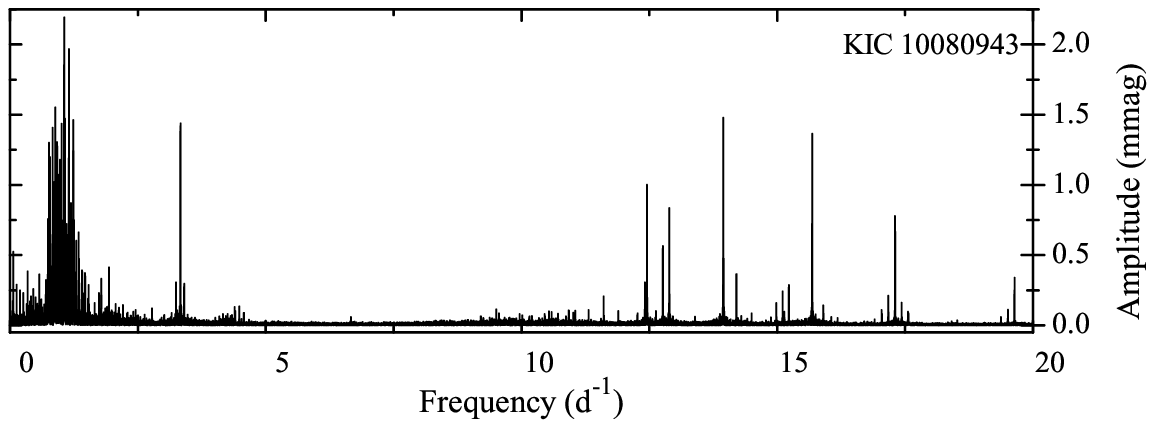}
\caption{\small Light curves (left) and amplitude spectra (right) of
selected stars. From top to bottom: KIC\,02710594, 03448365,
03744571, 04547348, 04749989, and 10080943. The y-scale of the left
panel represents differential magnitude after subtraction of the
corresponding mean value for each target. Note mmag scale for
KIC\,04749989.} \label{Figure: Selected_LC_and_AS}
\end{figure*}

For the analysis, we use all data that was publicly available by the
time of the analysis (Q0-Q8) and our $Kepler$ Guest Observer data
\citep[GO,][]{Still2011}. LC data are well-suited for our analysis
as the corresponding Nyquist frequency of 24.47~\cd\ (283.12~\mhz)
is well above the \GD-typical oscillation periods that lie between
0.5 and 3 days. Instead of using the standard pre-extracted light
curves delivered through MAST (Mikulski Archive for Space
Telescopes), we have extracted the light curves from the pixel data
information using custom masks based on software developed by one of
us (SB, Bloemen et al., in prep). Contrary to the standard masks,
which only use pixels with the highest signal-to-noise (S/N) ratios,
our masks contain as many pixels as possible with significant flux.
The light curves obtained by including these lower S/N pixels have
significantly less instrumental trends than the standard light
curves. Table~\ref{Table: Mask comparison} shows the influence of
the proper mask choice on the frequency analysis of one of our \GD\
stars. The fifteen crossed out low frequencies present in the
standard pre-extracted light curve are gone from the data when we
extract the light curve from the pixel data using a custom mask. An
appropriate removal of the instrumental effects is essential for
\GD\ stars pulsating in high-order g-modes that show up in the
low-frequency domain. One must thus cautiously interpret frequencies
deduced from the standard light curves \citep[as listed in,
e.g.][]{Grigahcene2010,Uytterhoeven2011,Balona2011b}.

We acquired high-resolution spectroscopy for the brightest targets
with the HERMES \citep[High Efficiency and Resolution Mercator
Echelle Spectrograph,][]{Raskin2011} spectrograph attached to the
1.2-m Mercator telescope (La Palma, Canary Islands). The spectra
have a resolution of 85\,000 and cover the wavelength range from 377
to 900~nm. Typical S/N ratios measured at $\lambda\lambda$~5500~\AA\
are about 40 for each spectrum. An overview of spectroscopic
observations is given in Table~\ref{Table: Observations}.

The spectra have been reduced using a dedicated pipeline. The data
reduction procedure included bias and stray-light subtraction,
cosmic rays filtering, flat fielding, wavelength calibration by
ThArNe lamp, and order merging. The normalisation to the local
continuum has been performed in two steps: first, by fitting a
``master function'' (cubic spline connected through some tens of
carefully selected continuum points) that accounts for the
artificial spectral signature induced by the flat field procedure.
The spectral signature of the flat field lamps is attenuated by red
blocking filters \citep{Raskin2011} which induces a specific
spectral signature in the reduced merged spectra. And second, a
correction for small-scale effects by selecting more ``knot points''
at wavelengths free of spectral lines and connecting them with
linear functions. At this phase, all spectra are corrected for their
individual radial velocities (RVs) as the position of the selected
``knot points'' is fixed in wavelength. For more information on the
normalisation procedure we refer to \citet{Papics2012b}.

\section{The target sample}\label{Section: The target_sample}

Our selection of \GD\ targets is based on an automated supervised
classification method, applied to the entire $Kepler$ Q1 dataset of
about 150\,000 light curves. For detailed information on the method,
we refer to \citet{Debosscher2011}. Basically, we summarize the main
characteristics of each light curve by means of a uniform set of
parameters, obtained from a Fourier decomposition. These include,
e.g., significant frequencies present in the light curve, their
amplitudes and their relative phases. Statistical class assignment
is then performed by comparing these parameters with those derived
from a training set of known class members for several types of
variable stars. This classification procedure only uses information
derived from the $Kepler$ light curve, not sufficient to distinguish
all variability types. Therefore, we refined the classification in a
second stage by using 2MASS colour information. This way, we could
discriminate between pulsating stars showing the same type of
variability in their light curves, but residing at different
locations in the HR diagram (e.g. \GD\ versus SPB pulsators).

In total, we selected 69 \GD\ candidate stars. Figure~\ref{Figure:
Sample representation} represents the entire sample by showing all
frequencies we consider as independent ones (cf.
Section~\ref{Section: Frequency analysis}) for all 69 targets. The
total number of shown frequencies is 344, which results from about 5
individual peaks per object. Such low number of independent modes is
plain statistical and does not mean that there are only 5
independent, non combination frequencies per star. In any
sufficiently large set of frequencies, it is not expected to find
more than $\sim$10 (depending on the criteria used to search for
combination frequencies) independent modes to occur as there is no
way to distinguish for the high-order combination and lower
amplitude modes. Most of the frequencies in Fig.~\ref{Figure: Sample
representation} are clearly grouped in the low-frequency region
below 3.5~\cd\ (40.50~\mhz). There is also a clear contribution in
the high-frequency domain. This points to the fact that \GD/\DSct\
hybrid pulsators have been included into the sample as well.

Figure~\ref{Figure: Selected_LC_and_AS} illustrates $Kepler$ light
curves and amplitude spectra of six selected stars: KIC\,02710594,
03448365, 03744571, 04547348, 04749989, and 10080943. The first four
targets belong to the same class of \GD-type variable stars and
exhibit similar variability in their light curves, though with
somewhat different beating patterns caused by the closely spaced
individual peaks in the low frequency domain. A much higher
frequency contribution is easily recognizable in the light curves of
the two other stars, KIC\,04749989 and 10080943, suggesting the
hybrid nature of these objects. Indeed, lots of high and moderate
amplitude peaks characteristic of p-mode pulsations show up in the
high frequency domain in the corresponding amplitude spectra of
these stars.

Table~\ref{Table: Photometric_class} presents photometric
characterization and classification of all 69 stars in our sample.
The first column gives the KIC-number of the star, the second and
the third columns represent the total frequency range in \cd\ and
\mhz, accordingly. In the next three columns we give the amplitude
range in mmag and frequency of the highest amplitude peak both in
\cd\ and \mhz, accordingly. The total number of statistically
significant frequencies as well as the photometric classification
according to the type of variability are given in the last two
columns. Despite our advanced data reduction procedure (cf.
Secion~\ref{Section: Observations}), for the majority of the stars,
low-frequency ($<$0.1~\cd\ or 1.2~\mhz) individual contributions
have been detected. As stated by \citet{Balona2011a}, the amplitudes
of such low frequencies are affected by the instrumental effects,
even in the case where the frequencies represent oscillation modes
of the stars. We are aware of that and thus do not take these
low-frequency contributions into account when classifying the
targets. Assignment of a star to a certain variability class (\GD\
or hybrid pulsator) has been made also by taking the information on
combination frequencies into account (cf. Section~\ref{Section:
Frequency analysis}). As such, the detection of individual peaks in
the high frequency domain is not by itself a sufficient condition
for the star to be classified as a \GD/\DSct\ hybrid pulsator -- the
peaks in the p-mode regime must be independent modes and not
(low-order) combinations of the low frequency g-modes. There are
several objects marked with superscript ``*'' in Table~\ref{Table:
Photometric_class} for which classification is uncertain in a sense
that some low-amplitude, high-order (typically, higher than 6-8)
combination frequencies have been detected in the high frequency
domain as well. Such high-order combinations, however, can be a
plain mathematical coincidence and we thus put some ``uncertainty
mark'' on these targets and are aware of the fact that they can also
exhibit very low-amplitude p-modes.

\begin{sidewaystable*}\vspace{180mm}
\tiny \centering \tabcolsep 1.9mm\caption{\small Photometric
classification and characterization of the sample targets. The stars
are sorted according to the KIC number.}\label{Table:
Photometric_class}
\begin{tabular}{llllllll|llllllll}
\hline\hline
\multicolumn{1}{c}{KIC}\rule{0pt}{9pt} & \multicolumn{2}{c}{Freq range} & \multicolumn{1}{c}{Ampl range} & \multicolumn{2}{c}{Freq$^{\rm high}$} & \multicolumn{1}{c}{N} & \multicolumn{1}{c}{Class} & \multicolumn{1}{|c}{KIC}\rule{0pt}{9pt} & \multicolumn{2}{c}{Freq range} & \multicolumn{1}{c}{Ampl range} & \multicolumn{2}{c}{Freq$^{\rm high}$} & \multicolumn{1}{c}{N} & \multicolumn{1}{c}{Class}\\
 & \multicolumn{1}{c}{\cd} & \multicolumn{1}{c}{\mhz} & \multicolumn{1}{c}{mmag} & \multicolumn{1}{c}{\cd} & \multicolumn{1}{c}{\mhz} & & & & \multicolumn{1}{c}{\cd} & \multicolumn{1}{c}{\mhz} & \multicolumn{1}{c}{mmag} & \multicolumn{1}{c}{\cd} & \multicolumn{1}{c}{\mhz} & &\\
\hline 02710594\rule{0pt}{9pt} & [0.03:3.00] & [0.35:34.73] &
[0.14:6.14] &  1.3553 & 15.6810 & 157 & \GD & 07746984 & [0.01:4.10]
& [0.12:47.42] & [0.03:2.51] &  1.3517 & 15.6394 & 109
& \GD\\
03222854 & [0.01:5.87] & [0.12:67.89] & [0.05:4.16] &  1.0575 & 12.2352 & 194 & \GD & 07867348 & [0.01:13.99] & [0.12:161.90] & [0.03:2.16] &  1.1699 & 13.5360 & 166 & \GD/\DSct\ hybrid\\
03448365 & [0.01:3.46] & [0.12:40.08] & [0.19:7.44] &  1.5001 &
17.3565 & 148 & \GD & 07939065 & [0.03:6.29] & [0.35:72.76] & [0.13:10.35] &  1.7282 & 19.9948 & 161 & \GD$^*$\\
03744571 & [0.01:2.90] & [0.12:33.59] & [0.05:7.86] & 0.9912 &
11.4685 & 204 & \GD & 08364249 & [0.01:5.89] & [0.12:68.09] & [0.01:3.86] &  1.8693 & 21.6282 & 198 & \GD\\
03952623 & [0.01:24.47] & [0.12:283.11] & [0.03:2.29] &  2.5707 &
29.7429 & 218 & \GD/\DSct\ hybrid & 08375138 & [0.01:6.31] & [0.12:73.06] & [0.03:5.12] &  2.0778 & 24.0402 & 379 & \GD$^*$\\
04547348 & [0.02:2.62] & [0.23:30.30] & [0.13:8.44] &  1.2249 &
14.1725 & 178 & \GD & 08378079 & [0.01:1.88] & [0.12:21.80] & [0.01:1.59] &  0.5254 & 6.0789 & 220 & \GD\\
04749989 & [0.02:22.76] & [0.23:263.36] & [0.01:0.52] &  3.1157 &
36.0492 & 115 & \GD/\DSct\ hybrid & 08611423 & [0.02:2.48] & [0.23:28.65] & [0.11:2.99] &  0.8297 & 9.5993 & 98 & \GD\\
04757184 & [0.02:5.08] & [0.23:58.73] & [0.02:11.30] &  1.2647 &
14.6327 & 153 & \GD & 08645874 & [0.01:24.25] & [0.12:280.55] & [0.01:5.58] &  1.8476 & 21.3769 & 248 & \GD/\DSct\ hybrid\\
04846809 & [0.01:2.27] & [0.12:26.21] & [0.11:2.75] &  1.8131 &
20.9780 & 105 & \GD & 08693972 & [0.01:1.82] & [0.12:21.09] & [0.02:5.81] &  0.4833 & 5.5923 & 381 & \GD\\
05114382 & [0.03:3.75] & [0.35:43.35] & [0.03:3.66] & 0.9542 &
11.0402 & 99 & \GD & 08739181 & [0.01:3.25] & [0.12:37.62] & [0.02:18.00] &  0.8605 & 9.9560 & 202 & \GD\\
05254203 & [0.02:4.45] & [0.23:51.52] & [0.05:9.33] &  1.2572 &
14.5460 & 166 & \GD & 08836473 & [0.01:9.50] & [0.12:109.91] & [0.02:1.44] &  1.8834 & 21.7913 & 217 & \GD/\DSct\ hybrid\\
05350598 & [0.02:4.49] & [0.23:51.94] & [0.02:0.90] &  2.1585 &
24.9740 & 139 & \GD & 09210943 & [0.02:5.29] & [0.23:61.25] & [0.03:2.36] &  2.1908 & 25.3475 & 176 & \GD\\
05522154 & [0.01:18.61] & [0.12:215.36] & [0.02:1.56] &  3.0101 &
34.8269 & 103 & \GD/\DSct\ hybrid & 09419694 & [0.06:2.86] & [0.69:33.10] & [0.21:14.34] &  1.0328 & 11.9498 & 122 & \GD\\
05708550 & [0.01:2.70] & [0.12:31.21] & [0.06:2.71] &  1.1155 &
12.9060 & 209 & \GD & 09480469 & [0.01:4.58] & [0.12:52.97] & [0.10:7.00] &  1.9948 & 23.0801 & 205 & \GD\\
05772452 & [0.01:2.11] & [0.12:24.43] & [0.20:11.96] &  0.7039 &
8.1440 & 96 & \GD & 09595743 & [0.01:3.54] & [0.12:40.89] & [0.13:7.04] &  1.7299 & 20.0146 & 96 & \GD\\
05788623 & [0.01:1.86] & [0.12:21.48] & [0.43:9.57] &  0.7790 &
9.0130 & 114 & \GD & 09751996 & [0.04:7.01] & [0.46:81.16] & [0.08:2.43] &  1.2846 & 14.8629 & 81 & \GD$^*$\\
06185513 & [0.03:9.76] & [0.35:112.93] & [0.01:0.69] &  2.2458 &
25.9843 & 145 & \GD/\DSct\ hybrid & 10080943 & [0.05:21.07] & [0.58:243.78] & [0.04:1.99] &  1.0593 & 12.2563 &257 & \GD/\DSct\ hybrid\\
06342398 & [0.01:2.67] & [0.12:30.85] & [0.04:7.27] &  1.0745 &
12.4325 & 145 & \GD & 10224094 & [0.01:3.02] & [0.12:34.95] & [0.04:2.78] &  1.0124 & 11.7131 &160 & \GD\\
06367159 & [0.02:19.07] & [0.23:220.67] & [0.12:3.92] &  0.5247 &
6.0710 & 177 & \GD/\DSct\ hybrid & 10256787 & [0.01:3.10] & [0.12:35.90] & [0.05:6.71] &  1.0775 & 12.4667 & 257 & \GD\\
06425437 & [0.03:2.21] & [0.35:25.52] & [0.54:15.63] &  0.8968 &
10.3756 & 80 & \GD & 10467146 & [0.01:3.82] & [0.12:44.15] & [0.02:3.42] &  0.9550 & 11.0489 & 253 & \GD\\
06467639 & [0.01:22.39] & [0.12:259.03] & [0.03:3.40] &  1.7361 &
20.0868 & 197 & \GD/\DSct\ hybrid & 11080103 & [0.02:4.17] & [0.23:48.20] & [0.08:11.97] &  1.2414 & 14.3632 & 170 & \GD\\
06468146 & [0.02:14.93] & [0.23:172.71] & [0.03:1.17] &  1.5457 &
17.8839 & 160 & \GD/\DSct\ hybrid & 11099031 & [0.01:11.93] & [0.12:137.98] & [0.01:1.51] &  0.9182 & 10.6241 & 325 & \GD$^*$\\
06468987 & [0.01:9.83] & [0.12:113.78] & [0.02:4.45] &  1.9990 &
23.1288 & 275 & \GD$^*$ & 11196370 & [0.01:6.14] & [0.12:71.08] & [0.03:2.71] &  2.8416 & 32.8769 & 275 & \GD$^*$\\
06678174 & [0.05:4.97] & [0.58:57.49] & [0.01:3.17] &  1.1283 &
13.0547 & 149 & \GD & 11294808 & [0.01:6.94] & [0.12:80.34] & [0.01:2.34] &  2.2248 & 25.7408 & 344 & \GD$^*$\\
06778063 & [0.01:23.81] & [0.12:275.50] & [0.04:0.93] &  18.0070 &
208.3414 & 164 & \DSct/\GD\ hybrid & 11456474 & [0.01:4.34] & [0.12:50.21] & [0.03:4.12] &  1.4715 & 17.0251 & 273 & \GD\\
06935014 & [0.01:2.67] & [0.12:30.85] & [0.04:5.47] &  1.2067 &
13.9621 & 333 & \GD & 11668783 & [0.01:4.79] & [0.12:55.38] & [0.01:3.13] &  0.6280 & 7.2665 & 225 & \GD\\
06953103 & [0.02:3.85] & [0.23:44.49] & [0.77:33.79] &  1.2876 &
14.8971 & 125 & \GD & 11721304 & [0.01:1.94] & [0.12:22.47] & [0.11:4.35] &  0.7905 & 9.1459 & 239 & \GD\\
07023122 & [0.01:5.86] & [0.12:67.79] & [0.04:13.02] &  1.8760 &
21.7058 & 259 & \GD & 11754232 & [0.01:8.19] & [0.12:94.70] & [0.03:2.13] &  0.9196 & 10.6396 & 129 & \GD/\DSct\ hybrid\\
07365537 & [0.01:18.40] & [0.12:212.86] & [0.01:7.77] &  2.9257 &
33.8502 & 399 & \GD$^*$ & 11826272 & [0.01:2.76] & [0.12:31.96] & [0.06:11.93] &  0.8337 & 9.6456 & 214 & \GD\\
07380501 & [0.01:4.14] & [0.12:47.85] & [0.02:2.27] &  0.9633 &
11.1449 & 279 & \GD & 11907454 & [0.01:4.33] & [0.12:50.10] & [0.05:4.35] &  1.1872 & 13.7358 & 246 & \GD\\
07434470 & [0.01:14.42] & [0.12:166.83] & [0.01:3.79] &  1.6987 &
19.6541 & 259 & \GD/\DSct\ hybrid & 11917550 & [0.01:3.07] & [0.12:35.53] & [0.08:9.18] &  1.2877 & 14.8985 & 219 & \GD\\
07516703 & [0.02:3.65] & [0.23:42.22] & [0.03:1.43] &  1.8271 &
21.1396 & 252 & \GD & 11920505 & [0.01:2.83] & [0.12:32.77] & [0.25:13.03] &  1.1988 & 13.8700 & 162 & \GD\\
07583663 & [0.01:4.62] & [0.12:53.49] & [0.06:6.28] &  1.0448 &
12.0879 & 242 & \GD & 12066947 & [0.01:6.68] & [0.12:77.23] & [0.01:2.96] &  2.7243 & 31.5198 & 226 & \GD$^*$\\
07691618 & [0.01:2.00] & [0.12:23.14] & [0.03:2.63] &  0.8219 &
9.5089 & 251 & \GD & 12458189 & [0.01:8.05] & [0.12:93.15] & [0.02:4.31] &  1.0396 & 12.0287 & 294 & \GD$^*$\\
 & & & & & & & & 12643786 & [0.01:4.88] & [0.12:56.44] & [0.13:16.86] &  1.6129 & 18.6612 & 185 & \GD\\
 \hline \multicolumn{16}{l}{$^*$possibly \GD/\DSct\
hybrid\rule{0pt}{9pt}}
\end{tabular}
\end{sidewaystable*}

\section{Frequency analysis of the $Kepler$ light curves}\label{Section: Frequency analysis}

\begin{table}
\caption{Fundamental stellar parameters. N gives the number of
individual spectra, error bars are 1-$\sigma$ confidence
level.}\label{Table: Fundamental parameters}
\begin{tabular}{llllll}
\hline\hline
\multicolumn{1}{c}{KIC\rule{0pt}{9pt}} & \multicolumn{1}{c}{\te} & \multicolumn{1}{c}{\logg} & \multicolumn{1}{l}{\vsini} & \multicolumn{1}{c}{[M/H]} & \multicolumn{1}{c}{N}\\
 & \multicolumn{1}{c}{K} & \multicolumn{1}{c}{dex} & \kms & \multicolumn{1}{c}{dex} &\\
\hline
02710594\rule{0pt}{11pt} & 6830$^{+155}_{-155}$ & 3.55$^{+0.55}_{-0.55}$ & 76.0$^{+5.5}_{-5.5}$ & --0.22$^{+0.27}_{-0.27}$ & 4\vspace{1.5mm} \\
03448365 & 6975$^{+130}_{-130}$ & 4.00$^{+0.35}_{-0.35}$ & 88.0$^{+5.5}_{-5.5}$ & --0.03$^{+0.15}_{-0.15}$ & 2\vspace{1.5mm}\\
04547348 & 7060$^{+130}_{-130}$ & 4.00$^{+0.50}_{-0.50}$ & 65.3$^{+5.0}_{-5.0}$ & --0.20$^{+0.15}_{-0.15}$ & 4\vspace{1.5mm}\\
04749989 & 7320$^{+120}_{-120}$ & 4.32$^{+0.35}_{-0.35}$ & 191.2$^{+10.5}_{-10.5}$ & +0.00$^{+0.12}_{-0.12}$ & 2\vspace{1.5mm}\\
04757184 & 7320$^{+130}_{-130}$ & 4.25$^{+0.50}_{-0.50}$ & 32.1$^{+4.2}_{-4.2}$ & --0.45$^{+0.20}_{-0.20}$ & 4\vspace{1.5mm}\\
05114382 & 7200$^{+110}_{-110}$ & 4.44$^{+0.40}_{-0.40}$ & 66.5$^{+5.0}_{-5.0}$ & --0.08$^{+0.20}_{-0.20}$ & 4\vspace{1.5mm}\\
05522154 & 7195$^{+100}_{-100}$ & 4.53$^{+0.35}_{-0.35}$ & 156.6$^{+12.0}_{-12.0}$ & --0.20$^{+0.20}_{-0.20}$ & 3\vspace{1.5mm}\\
05708550 & 7010$^{+110}_{-110}$ & 4.01$^{+0.35}_{-0.35}$ & 64.4$^{+4.5}_{-4.5}$ & --0.07$^{+0.14}_{-0.14}$ & 4\vspace{1.5mm}\\
06185513 & 7225$^{+180}_{-180}$ & 4.30$^{+0.75}_{-0.75}$ & 76.1$^{+12.0}_{-12.0}$ & --0.10$^{+0.30}_{-0.30}$ & 4\vspace{1.5mm}\\
06425437 & 7000$^{+170}_{-170}$ & 4.03$^{+0.57}_{-0.57}$ & 48.2$^{+5.5}_{-5.5}$ & +0.07$^{+0.22}_{-0.22}$ & 3\vspace{1.5mm}\\
06468146 & 7150$^{+105}_{-105}$ & 3.89$^{+0.40}_{-0.40}$ & 64.5$^{+4.0}_{-4.0}$ & +0.07$^{+0.14}_{-0.14}$ & 2\vspace{1.5mm}\\
06678174 & 7100$^{+105}_{-105}$ & 3.92$^{+0.40}_{-0.40}$ & 42.1$^{+3.5}_{-3.5}$ & --0.17$^{+0.14}_{-0.14}$ & 4\vspace{1.5mm}\\
06935014 & 7010$^{+100}_{-100}$ & 4.06$^{+0.30}_{-0.30}$ & 65.4$^{+4.0}_{-4.0}$ & +0.02$^{+0.12}_{-0.12}$ & 5\vspace{1.5mm}\\
07023122 & 7310$^{+105}_{-105}$ & 4.27$^{+0.31}_{-0.31}$ & 50.6$^{+3.4}_{-3.4}$ & --0.16$^{+0.15}_{-0.15}$ & 4\vspace{1.5mm}\\
07365537 & 7320$^{+90}_{-90}$ & 4.42$^{+0.31}_{-0.31}$ & 144.7$^{+9.0}_{-9.0}$ & --0.05$^{+0.15}_{-0.15}$ & 3\vspace{1.5mm}\\
07380501 & 6725$^{+145}_{-145}$ & 3.62$^{+0.65}_{-0.65}$ & 50.4$^{+4.5}_{-4.5}$ & --0.15$^{+0.20}_{-0.20}$ & 4\vspace{1.5mm}\\
07867348 & 6970$^{+120}_{-120}$ & 3.58$^{+0.40}_{-0.40}$ & 16.5$^{+2.7}_{-2.7}$ & --0.14$^{+0.13}_{-0.13}$ & 4\vspace{1.5mm}\\
08364249 & 6950$^{+135}_{-135}$ & 3.89$^{+0.60}_{-0.60}$ & 131.5$^{+12.0}_{-12.0}$ & --0.09$^{+0.17}_{-0.17}$ & 4\vspace{1.5mm}\\
08375138 & 7110$^{+100}_{-100}$ & 4.25$^{+0.33}_{-0.33}$ & 130.5$^{+8.0}_{-8.0}$ & --0.11$^{+0.10}_{-0.10}$ & 4\vspace{1.5mm}\\
08378079 & 7080$^{+180}_{-180}$ & 3.19$^{+0.70}_{-0.70}$ & 10.2$^{+1.4}_{-1.4}$ & --0.37$^{+0.20}_{-0.20}$ & 3\vspace{1.5mm}\\
08611423 & 7115$^{+155}_{-155}$ & 4.08$^{+0.55}_{-0.55}$ & 20.4$^{+3.2}_{-3.2}$ & --0.11$^{+0.13}_{-0.13}$ & 4\vspace{1.5mm}\\
08645874 & 7170$^{+105}_{-105}$ & 3.85$^{+0.35}_{-0.35}$ & 19.5$^{+1.6}_{-1.6}$ & --0.02$^{+0.10}_{-0.10}$ & 2\vspace{1.5mm}\\
09210943 & 7070$^{+130}_{-130}$ & 4.49$^{+0.50}_{-0.50}$ & 71.5$^{+8.0}_{-8.0}$ & --0.03$^{+0.20}_{-0.20}$ & 3\vspace{1.5mm}\\
09751996 & 6935$^{+135}_{-135}$ & 3.60$^{+0.45}_{-0.45}$ & 11.7$^{+1.5}_{-1.5}$ & --0.27$^{+0.12}_{-0.12}$ & 4\vspace{1.5mm}\\
10080943$^*$ & --- & --- & --- & --- & 4\vspace{1.5mm}\\
10224094 & 7030$^{+120}_{-120}$ & 3.82$^{+0.47}_{-0.47}$ & 23.7$^{+2.8}_{-2.8}$ & --0.07$^{+0.11}_{-0.11}$ & 4\vspace{1.5mm}\\
11099031 & 6795$^{+90}_{-90}$ & 3.97$^{+0.33}_{-0.33}$ & 96.5$^{+5.0}_{-5.0}$ & +0.12$^{+0.09}_{-0.09}$ & 4\vspace{1.5mm}\\
11294808 & 6875$^{+180}_{-180}$ & 3.93$^{+0.65}_{-0.65}$ & 87.8$^{+12.0}_{-12.0}$ & +0.07$^{+0.25}_{-0.25}$ & 4\vspace{1.5mm}\\
11721304 & 7195$^{+110}_{-110}$ & 4.25$^{+0.35}_{-0.35}$ & 26.5$^{+3.0}_{-3.0}$ & --0.08$^{+0.10}_{-0.10}$ & 4\vspace{1.5mm}\\
11826272 & 6945$^{+105}_{-105}$ & 3.79$^{+0.33}_{-0.33}$ & 28.2$^{+2.8}_{-2.8}$ & +0.00$^{+0.09}_{-0.09}$ & 3\vspace{1.5mm}\\
11907454 & 7040$^{+110}_{-110}$ & 4.27$^{+0.42}_{-0.42}$ & 105.6$^{+9.0}_{-9.0}$ & --0.03$^{+0.12}_{-0.12}$ & 4\vspace{1.5mm}\\
11917550 & 6990$^{+100}_{-100}$ & 3.93$^{+0.35}_{-0.35}$ & 74.0$^{+4.2}_{-4.2}$ & --0.09$^{+0.10}_{-0.09}$ & 4\vspace{1.5mm}\\
11920505 & 7100$^{+90}_{-90}$ & 4.04$^{+0.30}_{-0.30}$ & 59.5$^{+2.6}_{-2.6}$ & +0.01$^{+0.08}_{-0.08}$ & 2\vspace{1.5mm}\\
12066947 & 7395$^{+90}_{-90}$ & 4.59$^{+0.30}_{-0.30}$ & 122.5$^{+7.2}_{-7.2}$ & --0.12$^{+0.10}_{-0.10}$ & 3\vspace{1.5mm}\\
12458189 & 6895$^{+160}_{-160}$ & 3.99$^{+0.50}_{-0.50}$ & 67.5$^{+7.0}_{-7.0}$ & +0.08$^{+0.20}_{-0.20}$ & 4\vspace{1.5mm}\\
12643786 & 7160$^{+150}_{-150}$ & 4.20$^{+0.53}_{-0.53}$ & 72.5$^{+8.5}_{-8.5}$ & --0.19$^{+0.23}_{-0.23}$ & 4\vspace{1.5mm}\\
\hline \multicolumn{6}{l}{$^*$Spectroscopic double-lined binary
(SB2)\rule{0pt}{9pt}}
\end{tabular}
\end{table}

For the extraction of frequencies, amplitudes and phases of
pulsation modes from the $Kepler$ light curves, we used the
Lomb-Scargle periodogram \citep{Scargle1982} and a consecutive
prewhitening procedure. A detailed description of the whole
procedure can be found in, e.g., \citet{Papics2012b}.

In our case, at each step of the prewhitening procedure, we optimize
the amplitudes and phases of the modes by means of least-squares
fitting of the model to the observations while the frequencies fixed
to those obtained from the Scargle periodogram. This way, the
frequency uncertainty is determined by the frequency resolution
given by the Rayleigh limit 1/$T$, the errors in amplitudes and
phases are the formal errors from the least-squares fitting. We keep
iterating until the commonly used significance level of 4.0 in S/N
is reached \citep{Breger1993}, where the noise level is computed
from a 3~\cd\ (34.7~\mhz) window before prewhitening the frequency
peak of interest. However, there are several objects (typically
those for which less than three quarters of data are available) for
which ``only'' a couple of dozens of peaks can be detected following
this standard criterion, resulting in a fairly bad fit of the
observed light curve. In these few cases, we exceed the lower limit
in S/N and iterate until the value of S/N=3.0 is reached, making
sure the model fits the observations well.

\begin{figure*}[t]
\includegraphics[scale=0.98]{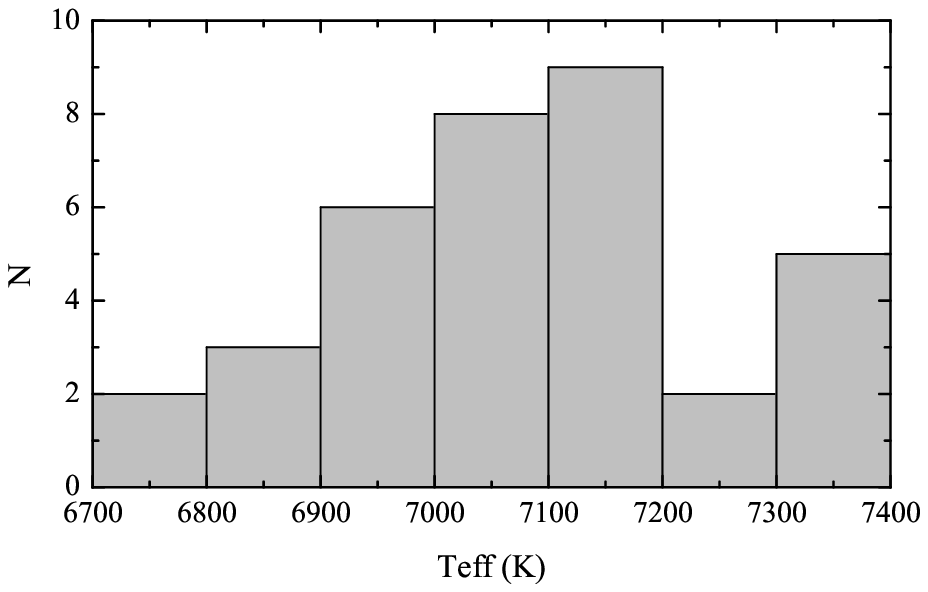}\hspace{2mm}
\includegraphics[scale=0.98]{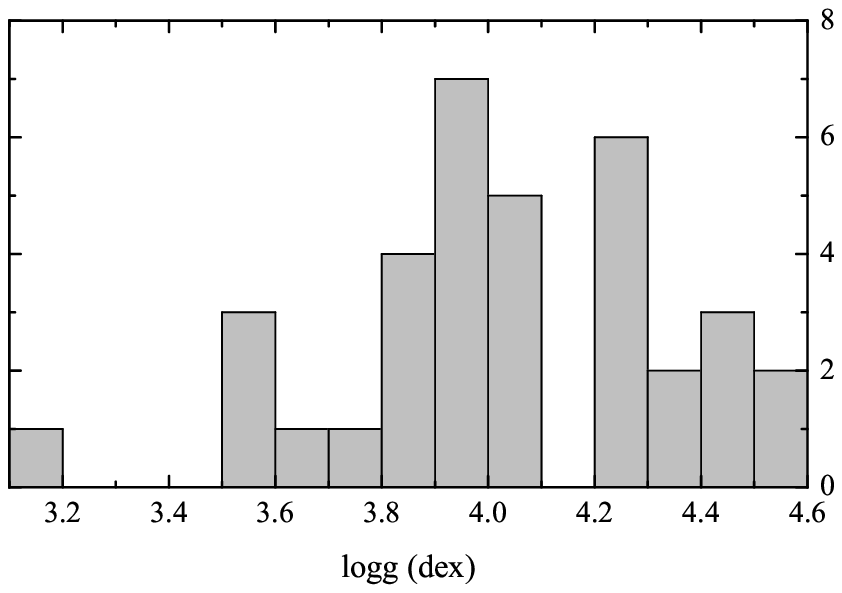}
\includegraphics[scale=0.98]{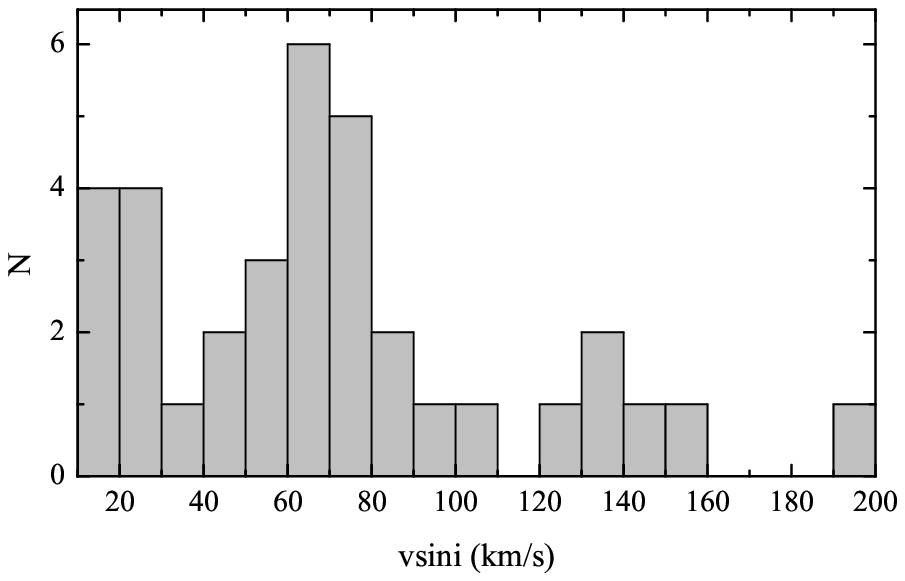}\hspace{10.5mm}
\includegraphics[scale=0.98]{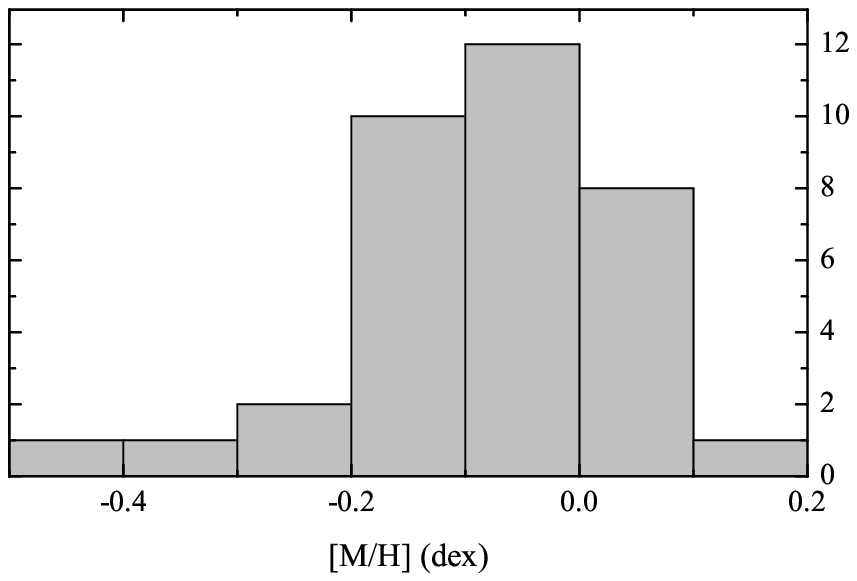}
\caption{\small Distribution of effective temperatures (top left),
surface gravities (top right), projected rotational velocities
(bottom left), and metallicities (bottom right) of the sample
targets.} \label{Histograms}
\end{figure*}

For every star in our sample, we check for evidence of non-linear
effects in the light curve by looking for low-order combination
frequencies. We assume that any combination frequency should have
lower amplitude than the parent frequencies, i.e. it should appear
in a list which is sorted according to amplitude, below the
frequency(ies) it is formed of. We allow up to three terms when
computing the sum and/or difference combination frequencies,
assuming that harmonics can enter the combination as well (e.g.,
f$_{\rm i}$=2f$_1$+f$_2$-f$_3$ or f$_{\rm i}$=f$_1$-3f$_2$+f$_3$).
The frequency of interest is accepted to be a combination of the
independent frequency if the difference between the combination and
true values is less or equal to the Rayleigh limit. For example, the
arbitrary frequency f$_{\rm i}$ is assumed to be a combination of
the independent frequencies f$_1$ and f$_2$ when, e.g.,
$\mid$f$_{\rm i}-$(2f$_1$+f$_2$)$\mid \leq$1/$T$.

Table~B.1 summarizes the results of the frequency analysis for all
69 stars. The total time span of the observations and thus Rayleigh
limit can vary from star to star and is indicated in the parenthesis
following the star designation. The table lists the frequencies
(both in \cd\ and \mhz) and corresponding amplitudes (in mmag), S/N,
the corresponding lowest order combination, and the number of
combinations that can be built of independent frequencies only. The
latter number obviously needs more explanation. Each frequency in
the list (except for the highest amplitude peak which is assumed to
be independent), is first considered to be a possible combination
term of the higher-amplitude parent frequencies. In the first step,
we check for all possible combinations of the parent frequencies
(all frequencies that have higher amplitudes than the one in
question) capable of representing within the Rayleigh limit the
frequency in question. Among the whole variety of the obtained
combinations (the number of which will be the larger the lower is
the amplitude of the frequency in question), we select those that
involve the independent frequencies only. The total number of such
combinations is given in the last column of Table~B.1 designated
``N''. A value of zero in this column means that the frequency of
interest can only be represented as a combination of the frequencies
which themselves are combinations of the independent peaks. In
practice, this means that the frequencies with the number N=0 can
only be represented as very high-order (typically, higher than 10)
combinations, which is basically just a mathematical coincidence and
has no physical meaning. The independent frequencies are highlighted
in Table~B.1 in boldface whereas their second- and third-order
combinations are shown in italics and with the superscript ``*'',
respectively.

\citet{Papics2012a} investigated the problem of occurring
combination frequencies in the light curves of $Kepler$ B-type
stars. The author simulated thousands of light curves having similar
but random power spectrum as the main-sequence SPBs.
\citet{Papics2012a} concluded that one has to restrict to the
low-order (up to the 3rd) combinations when interpreting the light
curves as the higher-order combination peaks are likely mere
coincidence than having any physical meaning. This is the reason why
we consider only low-order (2nd and 3rd) combination peaks as having
physical meaning while we are rather sceptical with respect to all
other, higher-order combinations.

\section{Spectroscopic analysis}\label{Section: Spectroscopic
analysis}

As mentioned in Sect.~\ref{Section: Observations}, for half of the
stars in our sample, we obtained high resolution (R=85\,000)
spectroscopic data. At least two spectra have been obtained for each
star in order to check for possible RV variations due to binarity.
Besides the binarity check, a collection of N independent spectra
allows to increase S/N by roughly a factor of $\sqrt{N}$ by
combining the single measurements and building an average spectrum.
The gain in S/N is very important given that we aim for estimation
of the fundamental stellar parameters like effective temperature
\te, surface gravity \logg, projected rotational velocity \vsini,
and metallicity [M/H].

As for the second half of the sample stars, their observations are
planned for the summer next year when the $Kepler$ field is best
visible on La Palma, Canary Islands. Besides that, we also plan more
extensive spectroscopic monitoring of the stars for which more than
just a couple of measurements would be essential (e.g.,
spectroscopic binaries for which a precise orbital solution and
possibly decomposed spectra in the case of double-lined binaries
(SB2) can be derived given that good orbital phase coverage is
provided). In the following, we describe the results of our spectrum
analysis of 36 of the sample stars and compare the spectroscopic
classification with the one based on the method by
\citet{Debosscher2011}.

\subsection{Fundamental parameters and position in the HR diagram}

For estimation of the fundamental parameters of the stars, we use
the GSSP code \citep{Tkachenko2012,Lehmann2011}. The code relies on
a comparison between observed and synthetic spectra computed in a
grid of \te, \logg, \vsini, [M/H], and microturbulent velocity
$\xi$, and finds the optimum values of these parameters from a
minimum in $\chi^2$. Besides that, individual abundances of
different chemical species can be adjusted in the second step
assuming a stellar atmosphere model of a certain global metallicity.
The grid of atmosphere models has been computed using the most
recent version of the LLmodels code \citep{Shulyak2004} in [--0.8,
+0.8]~dex, [4\,500, 22\,000]~K, and [2.5, 5.0]~dex range for
metallicity, effective temperature, and surface gravity,
respectively \citep{Tkachenko2012}. Synthetic spectra are computed
using the SynthV code \citep{Tsymbal1996} which allows to compute
the spectra based on individual elemental abundances, to take
vertical stratification of a chemical element and/or microturbulent
velocity into account.

The errors of measurement (1$\sigma$ confidence level) are
calculated from the $\chi^2$ statistics using the projections of the
hypersurface of the $\chi^2$ from all grid points of all parameters
onto the parameter in question. In this way, the estimated error
bars include any possible model-inherent correlations between the
parameters. Possible imperfections of the model like incorrect
atomic data, non-LTE effects, or continuum normalization are not
taken into account. In a recent study by
\citet{Molenda-Zakowicz2013} the use of different methods and codes
to derive atmospheric parameters for F, G, K, and M-type stars is
compared, and led the authors to conclude that the realistic
accuracy in the determination of atmospheric parameters for these
types of stars is $\pm$~150~K in \te, $\pm$~0.15~dex in [Fe/H], and
0.3~dex in \logg, even though error calculations for individual
programs might result in smaller errors. Hence, we are aware of a
possible underestimation of errors in Table~\ref{Table: Fundamental
parameters}.

\begin{figure}
\includegraphics[scale=0.9]{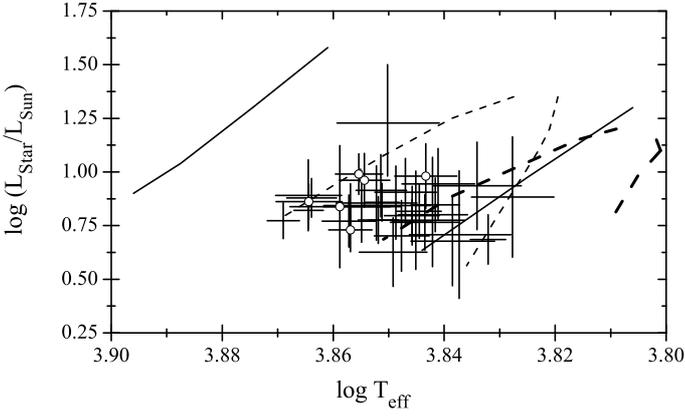}
\caption{\small Location of 36 stars (see Table~\ref{Table:
Fundamental parameters}) and the \GD\ and \DSct\ (solid lines)
theoretical instability strips \citep{Dupret2005} in the HR diagram.
Dashed thin and thick lines represent the \GD\ instability regions
computed with $\alpha=2.0$ and 1.5, respectively. Open circles refer
to the hybrid pulsators.} \label{Figure: HR diagram}
\end{figure}

Given the rather low S/N of our spectra (typically, below 80 for the
average spectrum), we decided to fix the microturbulent velocity
$\xi$ to the standard value of 2~\kms\ and optimise the global
metallicity (which is a scaling of all chemical elements heavier
than hydrogen and helium by the same factor) instead of evaluating
individual abundances of chemical elements. Table~\ref{Table:
Fundamental parameters} summarizes the results of the spectroscopic
analyses of the 36 stars in our sample listing the KIC-number, \te,
\logg, \vsini, and [M/H] values, as well as the number of individual
spectra used for building the average spectrum.

Figure~\ref{Histograms} shows distributions of the four fundamental
parameters of the sample stars. The majority of the targets (about
66~\%) have effective temperatures in the range between 6\,900 and
7\,200~K though a few objects show slightly higher/lower
($\pm$~200~K) temperature values. The distribution of the surface
gravities suggests that all but one stars are luminosity class IV-V
sub-giant or main-sequence stars, which together with the above
mentioned temperature regime, perfectly fits the classical
definition of the \GD\ stars group \citep[see,
e.g.][]{Kaye1999,Aerts2010}. The largest majority of the targets,
within the error of measurements, show solar metallicities though a
couple of stars exhibit slightly depleted overall chemical
composition compared to the Sun. Our sample is dominated by slowly
(\vsini$<$30~\kms) and moderately (30$<$\vsini$<$100~\kms) rotating
stars, only seven objects have \vsini\ values above 100~\kms.

\begin{figure}
\includegraphics[scale=0.88]{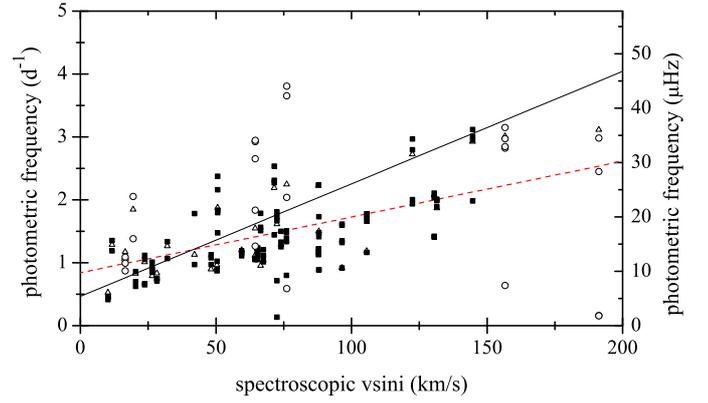}
\caption{\small \vsini\ versus independent frequencies for the
sample stars. Open triangles refer to the dominant oscillation mode
for each star, the two other symbols represent all other independent
frequencies. Filled boxes and open circles refer to \GD\ and hybrid
pulsators, respectively. The solid black line shows linear fit to
all data points (there are three frequencies between 18 (208.26) and
21 (242.97)~\cd\ (\mhz) in the \DSct\ domain), the red dashed line
represents a linear fit to the frequencies in the \GD\ range after
removing the \DSct\ frequencies.} \label{Figure: vsini_vs_freq}
\end{figure}

\begin{figure}
\includegraphics[scale=0.92]{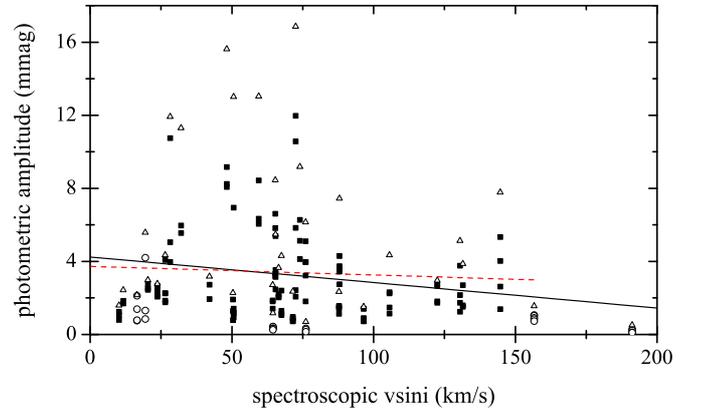}
\caption{\small Same as Fig.~\ref{Figure: vsini_vs_freq} but for the
amplitudes. The solid black line shows a linear fit to all data
points, the red dashed line represents a linear fit after removing
two stars with \vsini$>$150~\kms\ from the sample.} \label{Figure:
vsini_vs_ampl}
\end{figure}

The derived values of \te\ and \logg\ allow us to place the stars
into the log\,(L/L$_{\odot}$)--log\,\te\ diagram and classify them
by evaluating their positions relative to the \GD\ and \DSct\
theoretical instability strips. The luminosity of each star has been
estimated from an interpolation in the tables by
\citet{Schmidt-Kaler1982} using our spectroscopically derived \te\
and \logg. The errors in luminosity were evaluated by taking the
errors of both \te\ and \logg\ into account. However, we base our
spectroscopic classification mainly on the position of the stars
according to the derived temperatures as the luminosity errors can
still be underestimated due to the uncertainties in the empirical
relations.

Figure~\ref{Figure: HR diagram} shows the position of the stars in
the HR diagram together with the \GD\ and \DSct\ theoretical
instability strips. The latter are based on \citet[][Figures 2 and
9]{Dupret2005}. The edges of the \DSct\ instability region have been
computed for the fundamental mode and a mixing-length parameter of
$\alpha=1.8$ (solid lines). The borders of the \GD\ instability
regions computed with $\alpha=2.0$ and 1.5 are shown by dashed thin
and thick lines, respectively. All stars but one are nicely
clustering in the region of the HR diagram where the \GD-type g-mode
pulsations are expected to be excited in stellar interiors.
KIC\,08378079 is the only ``outlier'' located the top left (towards
higher temperatures and more luminous objects) of the \GD\
instability region in the diagram (cf. Figure~\ref{Figure: HR
diagram}). Given that our average spectrum is of very low S/N and
thus the error bars in both temperature and luminosity are large for
this faint ($V=12.5$) star, it can safely be considered as a
\GD-type variable from our spectroscopic values of \te\ and \logg.

As an overall conclusion, our spectroscopic classification perfectly
agrees and thus confirms the photometric one based on the
information derived from the $Kepler$ light curves only (cf.
Sect.~\ref{Section: The target_sample}). A second important
conclusion for future asteroseismic modelling is that most stars are
slow to moderate rotators. Of course, the estimated value of the
projected rotational velocity is not a measure of the true
rotational period of a star but allows to put some constraints
(lower limit) on the stellar rotation frequency.

\begin{figure}
\includegraphics[scale=0.84]{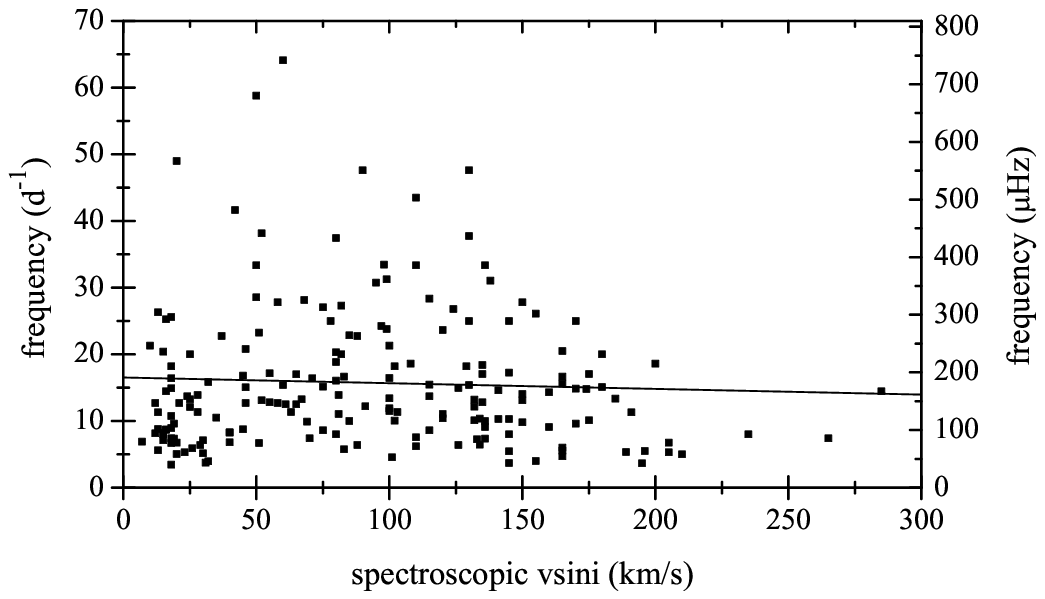}
\includegraphics[scale=0.84]{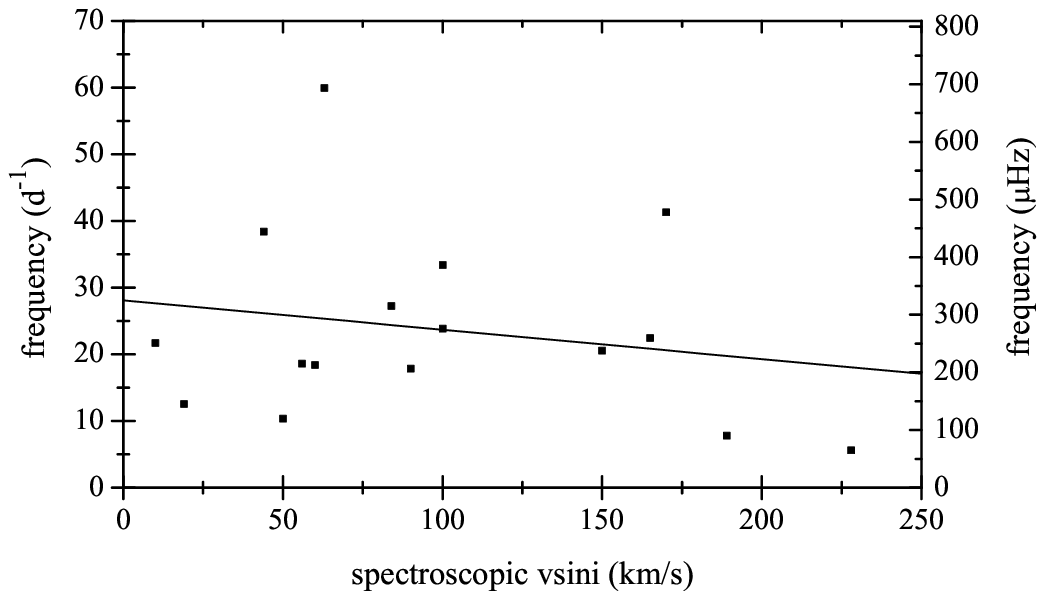}
\caption{\small Frequency of the dominant mode as a function of
\vsini\ for \DSct\ stars from the catalogues of
\citet[][top]{Rodriguez2000} and \citet[][bottom]{Uytterhoeven2011}.
The solid line represents a linear function fit to the shown data
points. Detailed information on individual stars is provided in
Table~A.1.} \label{Figure: vsini-frequency_DSct_R2000}
\end{figure}

We also checked for a possible correlation between the
spectroscopically derived value of \vsini\ and photometric
frequencies of the independent pulsation modes for each star. The
distribution is illustrated in Figure~\ref{Figure: vsini_vs_freq}.
There is a clear trend in the \vsini-frequency diagram implying that
the independent mode frequencies are higher for the larger \vsini.
The correlation becomes even stronger if we include the three
frequencies from the \DSct\ domain that we detected for two of the
shown stars (solid, black line to be compared with the dashed, red
line in Figure~\ref{Figure: vsini_vs_freq}). For completeness, in
Figure~\ref{Figure: vsini_vs_ampl}, we also present the
\vsini-amplitude distribution for the sample stars. Though a
negative trend shows up when fitting all the data points linearly
(solid line), it is almost gone when the two stars with
\vsini$>$150~\kms\ are removed (dashed line). Thus, we conclude that
the \vsini-amplitude correlation presented here is at best weak and
is not a characteristic of \GD\ stars.

To check whether a similar \vsini-frequency correlation is also
characteristic of \DSct\ stars, we used the data from the catalogues
of \citet[][hereafter called R2000]{Rodriguez2000} and
\citet{Uytterhoeven2011}. All stars for which \vsini\ measurements
are available were selected, resulting in 189 and 16 objects
extracted from R2000 and \citet{Uytterhoeven2011}, respectively. The
full list of the retrieved objects is given in Table~A.1. We are
also aware of the two recent extensive studies by
\citet{Balona2011c} and \citet{Chang2013} focusing on the \DSct\
stars but these authors do not present \vsini-frequency correlations
in their papers. The \vsini-frequency distributions obtained based
on the data from R2000 and \citet{Uytterhoeven2011} are illustrated
in Figure~\ref{Figure: vsini-frequency_DSct_R2000}. In both cases,
linear fits (solid lines in both panels) show a negative trend,
though with different slopes: it is steeper for the
\citet{Uytterhoeven2011} data than for those from R2000. However,
the reliability of the correlation shown in the bottom panel of
Figure~\ref{Figure: vsini-frequency_DSct_R2000} is not convincing
because of too few objects that could be retrieved from the sample
of \citet{Uytterhoeven2011}. Comparison of the R2000 distribution
(cf. Figure~\ref{Figure: vsini-frequency_DSct_R2000}, top panel)
with the one we obtained for the \GD\ stars in this paper (cf.
Figure~\ref{Figure: vsini_vs_freq}), reveals opposite behaviour: for
the \GD\ stars, the frequencies of oscillation modes are found to
increase as the \vsini\ increases whereas they decrease or, at most,
remain constant for the \DSct\ stars.

\begin{figure}
\includegraphics[scale=0.90]{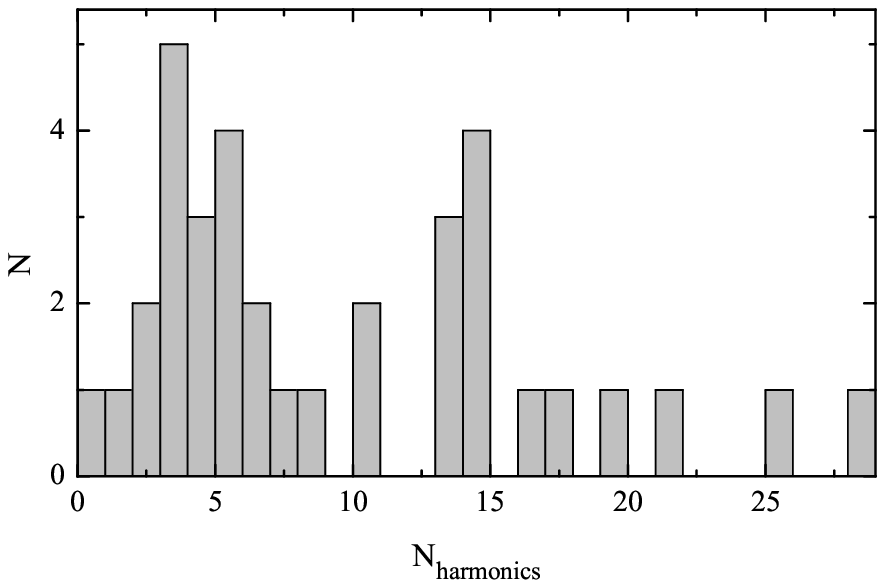}
\includegraphics[scale=0.90]{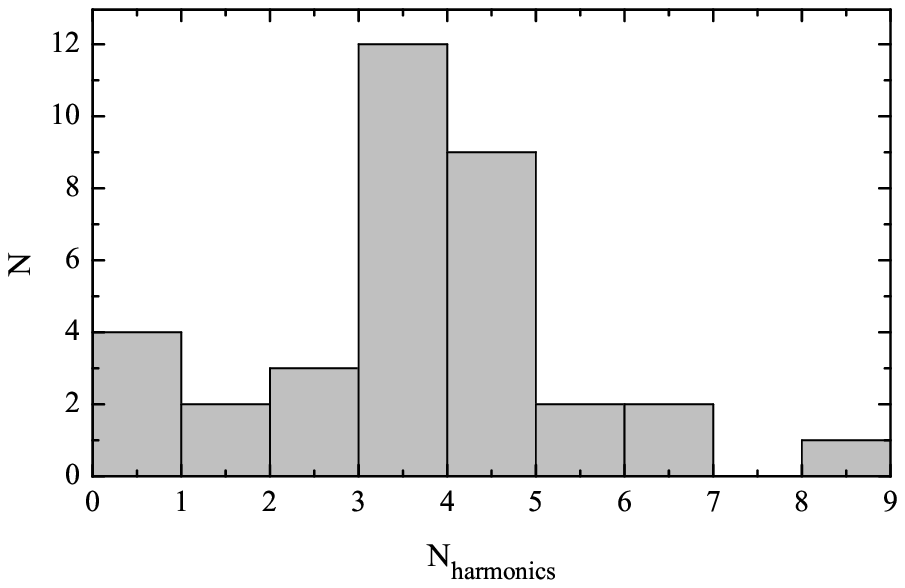}
\caption{\small Distributions of the total number of the detected
harmonics (top) and those formed of independent frequencies only
(bottom).} \label{Figure: histogram_harmonics}
\end{figure}

From Figure~\ref{Figure: vsini_vs_freq}, it is clear that not only
the dominant mode frequencies (open triangles) but also all the
other independent frequencies (filled boxes and open circles) show a
correlation with \vsini. If the correlation was due to rotational
modulation, one would expect only the dominant, rotation frequency
to show dependence on \vsini. Since we observe several independent
frequencies for each star and they all show similar behaviour with
respect to \vsini\ (cf. Figure~\ref{Figure: vsini_vs_freq}), we
conclude that the observed correlation is due to stellar pulsations
rather than rotational modulation. Moreover, besides the above
mentioned correlation for the rotation frequency, one would also
expect the rotational modulation to show up with series of harmonics
of that frequency \citep[see e.g.,][]{Thoul2013}. We thus checked
how many harmonics in total (including those of combination
frequencies) per star could be detected and present the distribution
in Figure~\ref{Figure: histogram_harmonics} (top). Though there is a
clear peak at three harmonics per star, the distribution is rather
random, revealing stars showing up to a couple of dozens of
harmonics in their light curves. The majority of these detections
are harmonics of (very low-amplitude) combination frequencies which
is likely to be just a mathematical coincidence rather than having
any physical meaning \citep[see][]{Papics2012a}. Indeed, none of the
stars show a series of harmonics but rather single peaks for
combination frequencies. We thus reconsidered our distribution
including only harmonics of the independent frequencies now (see
Figure~\ref{Figure: histogram_harmonics}, bottom). There is a clear
peak at three harmonics per star and the largest number of the
detected harmonics per object decreased from the previous 29 to the
present 9. A careful look at the independent frequencies and their
harmonics showed that the latter are homogeneously distributed among
all independent frequencies and there is no star showing a long
series of harmonics of a single independent frequency as one would
expect for rotational modulation.

\section{Discussion and conclusions}\label{Section: Conclusions}

Non-uniform period spacings of gravity pulsation modes is a powerful
tool for diagnostics of the properties of the inner core and its
surrounding layers. We aim for the application of the methodology
described by \citet{Miglio2008} and \citet{Bouabid2013} \citep[for
the first practical application, see][]{Degroote2010} to \GD-type
pulsating stars. This paper is the first step towards this goal.

Based on an automated supervised classification method
\citep{Debosscher2011} applied to the entire $Kepler$ Q1 dataset of
about 150\,000 high-quality light curves, we compiled a sample of 69
\GD\ candidate stars. We presented the results of frequency analyses
of the $Kepler$ light curves for all stars in our sample. For each
star, we checked the results for evidence of non-linear effects in
the light curve by looking for low-order combination frequencies.
All our stars show at least several second-order and third-order
combination frequencies (including harmonics) suggesting that
non-linear effects are common for \GD-type pulsating stars.

\begin{figure}[t]
\includegraphics[scale=0.96]{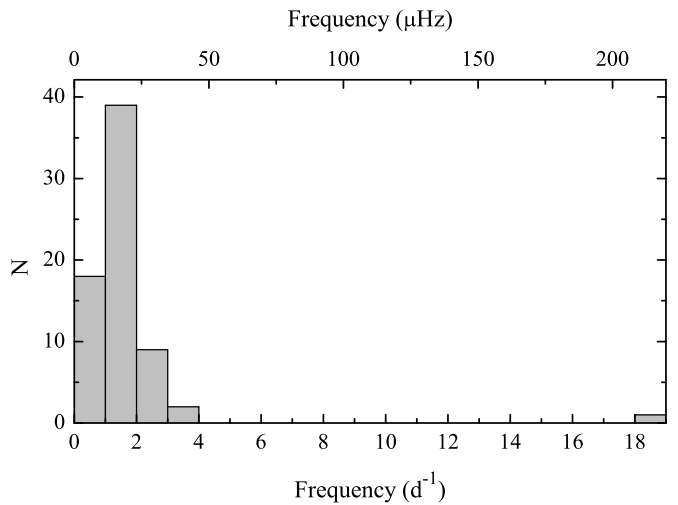}
\caption{\small Distribution of the highest amplitude frequency
detected in the Fourier spectra of the sample stars.}
\label{Histogram_frequency}
\end{figure}

From the $Kepler$ light curve analysis, we identified 45 \GD\ stars,
14 hybrid pulsators, and 10 stars marked as ``possible \GD/\DSct\
hybrids'' and showing low amplitude, high-order (typically, higher
than 6-8) combinations in the p-mode regime of the Fourier spectrum.
A more detailed study of the amplitude spectra would be needed for
these ten stars to classify them correctly. According to the highest
amplitude frequency distribution (cf.
Figure~\ref{Histogram_frequency}), the \GD-type g-mode pulsations
dominate our sample and there is only one star, KIC\,06778063, where
the peak in the p-mode regime is the dominant one in the data.

Additionally, high-resolution spectroscopic data have been obtained
for half of our stars with the HERMES spectrograph
\citep{Raskin2011} attached to the 1.2-m Mercator telescope. All 36
stars for which spectra have been acquired fall into the \GD\
instability region confirming the photometric classification as
either \GD- or of \GD/\DSct\ hybrid-type pulsating stars. The
effective temperatures are distributed within a narrow window of
700~K, the surface gravities distribution suggests luminosity class
IV-V sub-giant or main-sequence stars. Hence, the applied method of
photometric classification as described in \citet{Debosscher2011},
proves to be very robust.

\citet{Uytterhoeven2011} presented a characterization of a large
sample of A- to F-type stars based on $Kepler$ photometry and, where
available, ground-based spectroscopy. Among the stars showing \GD,
\DSct, or both types of pulsations in their light curves, the
authors identified about 36\% as hybrid pulsators. In our case, this
fraction is about 35\% if we consider 10 ``possible \GD/\DSct\
hybrids'' as being indeed hybrid pulsators and about 20\% if we
exclude them. An extrapolation of the results obtained by
\citet{Uytterhoeven2011} for a sub-sample of 41 targets to the
entire sample, led the authors to the conclusion that many of \GD\
and \DSct\ pulsators are moderate (40$<$\vsini$<$90~\kms) to fast
(\vsini$>$90~\kms) rotators. Our sample consists of mainly slow to
moderate rotators with the \vsini\ distribution peaking at the value
of 65~\kms\ (cf. Figure~\ref{Histograms}, bottom left). The major
conclusion that \citet{Uytterhoeven2011} made concerning \GD\ and
\DSct\ theoretical instability strips was that they have to be
refined as the stars with characteristic type of pulsations seem to
exist beyond the respective instability regions. The studies
preformed by \citet{Grigahcene2010,Tkachenko2012,Tkachenko2013} led
to similar conclusions. Also \citet{Hareter2012}, who studied a
large number of the CoRoT light curves, found that \GD\ stars mainly
cluster at the red edge of the \DSct\ instability strip, with a
significant fraction of them having cooler temperatures however,
whereas \DSct-\GD\ hybrid pulsators fill the whole \DSct\
instability region with a few stars being beyond its blue edge. The
author also concludes that, from the position of the studied objects
in the HR diagram, there is no close relation between \GD\ stars and
hybrid pulsators. In this paper, we found that all stars for which
fundamental parameters could be measured from spectroscopic data,
reside in the expected range of the HR diagram, clustering inside
the theoretical \GD\ instability region. Our results are based on
different i.e., much stricter, selection criteria than those applied
in the above mention papers, i.e. the morphology of the light curves
in combination with a spectroscopic \te\ determination rather than a
poor \te\ estimate or colour index alone. This implies that our
sample is much ``cleaner'' and not contaminated by stars with other
causes of variability than pulsations.

\citet{Balona2011b} suggested that the pulsation and rotation
periods must be closely related. Given that in
\citet{Uytterhoeven2011} \vsini\ values were obtained for a very
limited number of stars, the authors did not look for any
correlations between \vsini\ and pulsation period/frequency. In our
case, we have found a clear correlation between the
spectroscopically derived \vsini\ and the photometric frequencies of
the independent pulsation modes (cf. Figure~\ref{Figure:
vsini_vs_freq}) in a sense that the modes have higher frequencies
the faster is the rotation. This correlation is not caused by
rotational modulation of the analysed stars but is valid for the
\GD\ g-mode stellar oscillations. These findings are in perfect
agreement with the results of the recent theoretical work by
\citet{Bouabid2013}, who showed that the rotation shifts the g-mode
frequencies to higher values in the frequency spectrum of a
pulsating star, filling the gap between the \DSct-type p- and
\GD-type g-modes. Comparison with the \DSct\ stars reveals that they
behave in the opposite way, namely they show decreasing frequency as
the \vsini\ increases. We also checked for a \vsini-amplitude
correlation for our \GD\ stars but did not get any convincing
results.

Compared to the study by \citet{Uytterhoeven2011} where a large
sample of 750 A- to F-type stars has been analysed and of which
about 21\% were found not to belong the class of \GD\ or \DSct\ nor
to the hybrid pulsators, 12\% were identified as binary or multiple
star systems, and many objects were found to be constant stars, our
sample is much more homogeneous and does not include ``bias'', i.e.,
all selected stars are multiperiodic pulsators. Compared to the
sample of CoRoT stars compiled by \citet{Hareter2012}, nearly
continuous $Kepler$ observations spread over several years offer an
order of magnitude higher frequency resolution which allows to
resolve the frequency spectrum and makes the detection of frequency
and period spacings possible. Hence, our sample selection and
frequency analysis offer a good starting point for seismic modelling
of \GD\ stars.

In follow-up papers, we plan to present the detailed frequency
analysis results for all individual stars as well as a spectroscopic
analysis of the fainter stars in the sample, in addition to seismic
modelling based on the observational results.

\begin{acknowledgements}
The research leading to these results received funding from the
European Research Council under the European Community's Seventh
Framework Programme (FP7/2007--2013)/ERC grant agreement
n$^\circ$227224 (PROSPERITY). Funding for the {\it Kepler} mission
is provided by NASA's Science Mission Directorate. We thank the
whole team for the development and operations of the mission. This
research made use of the SIMBAD database, operated at CDS,
Strasbourg, France, and the SAO/NASA Astrophysics Data System. This
research has made use of the VizieR catalogue access tool, CDS,
Strasbourg, France.
\end{acknowledgements}

{}

\end{document}